\documentclass[12pt]{article}
\usepackage{graphicx}
\usepackage{amsmath}
\usepackage{amssymb}
\usepackage{amsthm}
\usepackage{ccaption}
\usepackage[usenames]{color}
\usepackage{cite}
\def\TL{\hfil$\displaystyle{##}$}
\def\TR{$\displaystyle{{}##}$\hfil}

\def\comment#1{}
\def\fixit#1{}

\def\overleftrightarrow#1{\vbox{\ialign{##\crcr
     $\leftrightarrow$\crcr\noalign{\kern-0pt\nointerlineskip}
     $\hfil\displaystyle{#1}\hfil$\crcr}}}

\def\lsim{\mathrel{\mathstrut\smash{\ooalign{\raise2.5pt\hbox{$<$}\cr\lower2.5pt\hbox{$\sim$}}}}}
\def\gsim{\mathrel{\mathstrut\smash{\ooalign{\raise2.5pt\hbox{$>$}\cr\lower2.5pt\hbox{$\sim$}}}}}

\def\sqr#1#2{{\vcenter{\vbox{\hrule height.#2pt
         \hbox{\vrule width.#2pt height#1pt \kern#1pt
            \vrule width.#2pt}
         \hrule height.#2pt}}}}

\def\href#1#2{#2}  

\def\lbldef#1#2{\expandafter\gdef\csname #1\endcsname {#2}}
\def\eqn#1#2{\lbldef{#1}{(\ref{#1})}%
\begin{equation} #2 \label{#1} \end{equation}}
\def\eqalign#1{\vcenter{\openup1\jot
    \halign{\strut\span\TL & \span\TR\cr #1 \cr
   }}}

\makeatletter
\@addtoreset{equation}{section}

\makeatletter
\renewcommand\section{\@startsection {section}{1}{\z@}%
                                   {-3.5ex \@plus -1ex \@minus -.2ex}
                                   {2.3ex \@plus.2ex}%
                                   {\normalfont\large\bfseries}}
\renewcommand\subsection{\@startsection{subsection}{2}{\z@}%
                                     {-3.25ex\@plus -1ex \@minus -.2ex}%
                                     {1.5ex \@plus .2ex}%
                                     {\normalfont\bfseries}}

\newcommand{\startappendix}{
\setcounter{section}{0}
\renewcommand{\thesection}{\Alph{section}}}

\newcommand{\Appendix}[1]{
\refstepcounter{section}
\vspace{10mm}
\pagebreak[3]
\setcounter{equation}{0}
\begin{flushleft}
{\large\bf Appendix \thesection: #1}
\end{flushleft}}

  \captionnamefont{\bfseries}
  \captiontitlefont{\small\sffamily}
  \captiondelim{: }
  \hangcaption

\def\baselinestretch{1.2}
\def\sec#1{Section \ref{#1}}
\def\fig#1{Fig.\,\ref{#1}}

\def\App#1{Appendix \ref{#1}}

\def\[{\left [}
\def\]{\right ]}
\def\({\left (}
\def\){\right )}

\def\eg{{\it e.g.}}
\def\vs{{\it vs.}}
\def\cf{{\it cf.}}
\def\ie{{\it i.e.}}

\def\etc{{\it etc.}}

\def\a{\alpha}

\def\Om{\Omega}

\def\ph{\varphi}
\def\d{\delta}
\def\D{\Delta}
\def\k{\kappa}
\def\eps{\varepsilon}

\def\thus{\Longrightarrow}
\def\thuss{\qquad \Longrightarrow \qquad}

\def\CN{{\cal N}}
\def\CM{{\cal M}}
\def\CO{{\cal O}}
\def\CS{{\cal S}}

\def\R{{\bf R}}
\def\S{{\bf S}}

\def\half{{\frac{1}{2}}}

\def\p{\partial}

\def\bra#1{\langle \, #1 \mid}
\def\ket#1{\mid \! #1 \, \rangle}

\def\expval#1{{\langle \, #1  \, \rangle}}

\def\abs#1{\mid \! #1 \! \mid}

\def\rb{{\bar r}}
\def\ro{\rho_0}

\def\ov{\over}
\def\ha{{1\over 2}}
\def\lam{{\lambda}}
\def\Lam{{\Lambda}}
\def\al{{\alpha}}
\def\ket#1{|#1\rangle}
\def\bra#1{\langle#1|}
\def\vev#1{\langle#1\rangle}

\def\Om{{\Omega}}

\def \lam {\lambda}

\def \ga {\gamma}
\def\sig{{\sigma}}
\def\ep{{\epsilon}}

\def\OO{{\cal O}}

\def\GG{{\cal G}}

\newcommand{\be}{\begin{equation}}
\newcommand{\ee}{\end{equation}}
\newcommand{\bea}{\begin{eqnarray}}
\newcommand{\eea}{\end{eqnarray}}

\def\bM{\partial \CM}
\def\rh{r_{+}}
\def\SAdS{Schwarzschild-AdS}
\def\I{${\cal{A}}$}
\def\II{${\cal{B}}$}
\def\am{\a_0}

\title{{\bf Bulk-cone singularities \& signatures of horizon formation in AdS/CFT}}

\author{Veronika E Hubeny$^\dagger$, Hong Liu$^\sharp$, and Mukund Rangamani$^\dagger$
\footnote{veronika.hubeny@durham.ac.uk, hong\_liu@lns.mit.edu, mukund.rangamani@durham.ac.uk}\\ \\
\small \sl $^\dagger$ \; Centre for Particle Theory \& Department of
Mathematical Sciences,
\\[-1.5mm]
\small \sl Science Laboratories, South Road, Durham DH1 3LE, United Kingdom \\
\small \sl $^\sharp$ Center for Theoretical Physics , Massachusetts Institute of Technology,
\\[-1.5mm]
\small \sl Cambridge, Massachusetts, 02139, USA}

\begin{document}
\setlength{\baselineskip}{16pt}
\begin{titlepage}
\maketitle
\begin{picture}(10,10)(0,0)
\put(335,325){DCPT-06/29} 
\put(335,310){MIT-CTP-3775}
\end{picture}

\begin{abstract}
We discuss the relation between singularities of correlation
functions and causal properties of the bulk spacetime in the
context of the AdS/CFT correspondence. In particular, we argue
that the boundary field theory correlation functions are singular
when the insertion points are connected causally by a bulk null
geodesic. This implies the existence of ``bulk-cone
singularities'' in boundary theory correlation functions which lie
inside the boundary light-cone. We exhibit the pattern of singularities in various  asymptotically AdS spacetimes and argue that this
pattern can be used to probe the bulk geometry. We
apply this correspondence to the specific case of shell collapse
in AdS/CFT and indicate a sharp feature in the
boundary observables corresponding to black hole event horizon formation.
 \end{abstract}
\thispagestyle{empty}
\setcounter{page}{0}
\end{titlepage}

\renewcommand{\baselinestretch}{1.2}  

\section{Introduction}
\label{intro}

The AdS/CFT correspondence, which relates quantum gravity in asymptotically Anti de Sitter spacetimes to a non-gravitational gauge theory, is an invaluable window to
understanding effects of gravity beyond the semi-classical
approximation. In principle the correspondence provides a
framework to address many long standing questions in quantum
gravity, such as the nature of singularity resolution or issues
relating to the information paradox. While much has been learnt in
the past few years, the status of the
dictionary between gravitational quantities and field theory
observables is still at a somewhat rudimentary stage. This has been a
stumbling block for exploiting the correspondence to its full
power, especially in eliciting answers to oft posed quantum gravitational
questions.

In classical general relativity, concepts such as causal
structure, event horizons, singularities, \etc, play an important
role in understanding the geometry of the spacetime manifold. In
the semi-classical approximation these concepts are useful in
understanding the dynamics of quantum fields in curved
backgrounds. Given the AdS/CFT correspondence it is interesting to
understand field theoretic encoding of these geometric concepts.
Considering the central role played by geometry in classical general
relativity, one naively expects them to have a well defined
representation in the field theory.

For instance, consider the field theoretic representation of bulk
causal structure. In the semi-classical limit, the bulk causal
structure can be read off from the properties of correlation
functions of quantum fields propagating on the spacetime manifold.
For asymptotically AdS spacetimes the bulk correlation functions
in turn determine (in some suitable scaling limit) the boundary
correlation functions. Consistency within the correspondence
requires therefore that the boundary correlation functions are
compatible with the bulk causality constraints. This has been
demonstrated to be true in many examples
\cite{Horowitz:1999gf,Kabat:1999yq,Gao:2000ga,Hubeny:2005qu}. A
crucial ingredient in most of these analyses has been to exploit
the asymptotic AdS geometry where consistency is guaranteed by the
symmetries in question.

The non-trivial aspect of the geometry that one would like to
understand is the behaviour of the bulk causal structure deep in
the interior of an asymptotically AdS spacetime. For example one
can ask whether the presence of an event horizon in the spacetime
has a non-trivial signature in the boundary. The zeroth order
answer to this question, at least for eternal black holes  in
AdS spacetime, is that the dual field theory is in a thermal
state.\footnote{This is true only for black holes whose horizon
size is larger than the AdS scale.} So the  presence of the
horizon is encoded in a new scale for physics in the boundary
theory, the thermal scale. However, this information hardly probes
the causal structure in the vicinity of the horizon.

Furthermore, since the AdS/CFT correspondence is most firmly
understood in the Euclidean framework, questions pertaining to
genuinely time-dependent processes in the bulk and holographic
duals thereof become much more subtle. For example, consider 
modelling a time dependent process, such as  of black hole formation in the bulk,  within the field theory.   A more nuanced question, then, would to  ask whether there is any field theory information to be gained about the event horizon formation process.
Phrased differently:  {\it can we see horizon formation directly
in the gauge theory?}   Since in the bulk the event horizon forms at a
sharply-localized event\footnote{In ths work we consider spherically symmetric spacetimes. 
More generally, in non-spherical spacetimes the event horizon could form at a locus of events; 
our methods should generalise to these cases as well.}
 (despite the fact that we need to know the
full future evolution of the spacetime to actually find the
horizon), we would expect that this will manifest itself in some
correspondingly sharp feature in the gauge theory.

The conventional lore is that the UV/IR relation, which maps
local regions in the interior of the bulk to non-local objects in
the CFT, makes it hard to extract useful information about bulk
geometry (\cf, \cite{Gregory:2000an} for attempts to use UV/IR
relation to extract information about the horizon). Given that the
geometric scale associated with the horizon corresponds to the
thermal scale for neutral black holes,  it is hard to see how to extract a precise signal
about the causal structure in the neighbourhood of the event
horizon. 

The clue comes from the progress made in the recent past in
identifying the CFT signature of the black hole singularity
\cite{Fidkowski:2003nf,Festuccia:2005pi}.  In this case the field
theoretic observables transcend the classical barrier of the event
horizon to encode information of behind the horizon physics. The
main observation of \cite{Fidkowski:2003nf}, which built on the
original work of \cite{Louko:2000tp,Kraus:2002iv} was to use the
intrinsic non-locality of the boundary correlation functions to
identify signals of the singularity. The black hole singularity
was identified to correspond to a particular `light-cone' like
singularity in the field theory correlation function.\footnote{The
identification of the singularity in the strict large $N$ limit is
easiest when formulated in terms of momentum space correlators
\cite{Festuccia:2005pi} as opposed to direct computation in
position space where the singularity is not visible in the primary
sheet of the correlator \cite{Fidkowski:2003nf}. In particular, it
was confirmed in \cite{Festuccia:2005pi} that the signatures of the
singularities go away when the rank of the boundary theory gauge
group is finite, implying that the singularities are resolved in
quantum gravity.} It has further been proposed that this technique
can be exploited to understand physics behind the event horizon
and in particular used to investigate aspects of inflationary
geometry within AdS/CFT \cite{Freivogel:2005qh}. The basic idea
was to look at the bulk Green's functions in a saddle
approximation where they are dominated by geodesics and use the
bulk computations to derive predictions for boundary correlation
functions.

In the present work we argue that the horizon formation in gravitational collapse can be detected in the boundary theory by examining the structure of singularities of generic Lorentzian correlators. In particular,  we show that a sharp horizon-formation time can be extracted from the pattern of singularities.

Our argument is based on a connection between null geodesics in
the bulk spacetime and singularities in correlation functions of
local operators evaluated in the state corresponding to the bulk
geometry.\footnote{Spacetimes which are asymptotically AdS can be
thought of as deformations of pure AdS by normalizable modes in
the supergravity description. In field theory terms this
corresponds to a generic excited state obtained by acting on the
CFT vacuum with the appropriate operator.} The basic idea is that
CFT correlators will exhibit light-cone singularities when the
points of operator insertion are connected by a null geodesic. The
connection implies that CFT correlators in excited states have
additional Lorentzian singularities inside the light cone,
which we will call {\it bulk-cone singularities}. Usually CFT
correlators exhibit light-cone singularities and the location of
these is determined by the causal structure of the background
spacetime on which the field theory lives. For a CFT living on $\R
\times \S^{d-1}$ these would be given by the conventional light
cone of the Einstein static universe. Our main observation is that
bulk causality, together with the dictionary between bulk Green's function
and boundary correlation functions in the AdS/CFT context,  necessitate additional bulk-cone singularities in the CFT correlators.

Armed with this relation we can ask how certain geometric
structures are encoded in the boundary data. In addition to a
black hole collapse geometry which motivated this study, we also
look at the geometry of a star in AdS spacetime and the  eternal black hole geometry.  In all these cases, interesting details about the geometry and its causal structure
that can be read off from the properties of the singularities in
the CFT correlators. In fact, as has recently been shown in \cite{Hammersley:2006cp}, the location of the boundary singularities can be used to reconstruct the complete bulk metric for a class of static, spherically symmetric spacetimes.
Our analysis is carried out by explicitly
studying geodesics in the bulk spacetime and using this to infer
properties about correlation functions in strongly coupled gauge
theories. While there are a few consistency checks we can
establish in some simple cases, many of our results are
predictions for strong coupling CFT correlation functions in
certain excited states.

The plan of the paper is as follows: We begin in \sec{nullgsing}
by motivating our claim that Lorentzian AdS/CFT correspondence
implies that correlation functions of local operators evaluated in
excited states of the field theory exhibit light-cone
singularities whenever the operator insertion points are connected
by null geodesics through the bulk spacetime. We then confirm the
relation between light-cone singularities and bulk null geodesics
for vacuum correlators. Further, we generalize to excitations
about vacuum state and obtain predictions for the singularities in
the correlation functions from the behaviour of null geodesics. To demonstrate how sensitively the 
singularity pattern depends on the state, in Sections \ref{statgeom} and \ref{bhthermal} we explicitly study the nature of the singularities and their implications for correlation
functions in two scenarios: a generic excited state and a thermal state, which we
model in the bulk by a star in AdS and an eternal black hole
geometry, respectively. We devote \sec{horform} to our main example, the
collapse of a black hole in AdS spacetime, and elicit from the
geometry signatures of event horizon formation in the field
theory. We conclude in \sec{discuss} with a discussion. Finally,
in the Appendices we present more detailed calculations pertaining
to the various geometries considered  (pure AdS, star in AdS,
eternal \SAdS, Poincare patch, null shell collapse, and
Vaidya-AdS).

\section{Null geodesics and boundary singularities}
\label{nullgsing}

In this section we motivate a simple relation between singularities of the boundary correlation functions and null geodesics in the bulk spacetime. The main idea is to exploit the AdS/CFT
dictionary for evaluating strong coupling correlation functions in the CFT using bulk Green's functions. Since the bulk Green's functions are sensitive to the bulk causal structure we expect that the properties of the bulk light cone to be directly visible in the boundary correlation functions.

A short word on notation: for convenience, we will denote the $d+1$ dimensional bulk spacetime by $\CM$ and its $d$ dimensional boundary by $\bM$.  Further, we use labels $x, y$ \etc, for points on the boundary $\bM$ and $r$ to denote the radial coordinate normal to $\bM$.

\subsection{General argument}

The AdS/CFT correpsondence gives us a precise map relating string theory (or its low energy supergravity limit) on $\CM$ with a field theory living on the boundary $\bM$ of $\CM$. To be precise, in the low energy supergravity approximation of the bulk theory (which is good as long as we choose  the 't Hooft coupling $\lambda$ and the rank of the gauge group $N$ on the boundary to be large, \ie, $\lambda \gg 1$ and $N \gg 1$) we can obtain the boundary correlation function $G(x,x')$ from the bulk propagator $\GG (x,r;x';r')$ of the corresponding field  by taking both end points to the boundary. More precisely,
 \be \label{uroP}
 G(x,x') = 2\,  \nu \, \lim_{r \to \infty, r' \to \infty} \, (r
 r')^{\Delta} \, \GG (x,r;x';r') \ ,
 \ee
where $\GG$ is the free field theory propagator for the corresponding
bulk field. Focussing on the simple case of local gauge invariant operators $\CO(x)$
of conformal dimension  $\Delta$  in  $\bM$ which are dual to free scalar fields
parameterized by  mass $m$ in the bulk, in (\ref{uroP}) we have
 \be
 \Delta = {d \ov 2} + \nu \ , \qquad \nu = \sqrt{m^2 +{d^2 \ov 4}} \ ,
 \ee
where we have set the curvature radius of AdS to be $1$. Note that
by choosing the asymptotic AdS time to coincide with the boundary time,
one ensures that time ordering chosen for the bulk propagator $\GG$
carries over to the boundary correlator in the limiting procedure.
In what follows we will consider Feynmann propagators.

That the limit in (\ref{uroP}) is well-defined for an asymptotically
AdS spacetime can be seen as follows. Upon fixing $(x',r')$, $\GG
(x,r;x',r')$ is a normalizable solution of the free bulk wave
equation on $\CM$. This implies for an asymptotically AdS spacetime
 \be \label{lim1}
 \GG (x,r;x',r') \sim r^{-\Delta}, \qquad r \to
\infty  \ .
 \ee
Similarly,
  \be \label{lim2}
 \GG (x,r;x',r') \sim r'^{-\Delta}, \qquad r' \to
\infty  \ .
 \ee
 Thus for generic boundary points $x$
and $x'$, the limit (\ref{uroP}) is well defined and yields a
regular $G(x,x')$. However, if $x$ and $x'$ are connected by a
null geodesic, (\ref{lim1}) and (\ref{lim2}) could break down, in
which case one expects 
 \be \label{oep}
 \lim_{r,r' \to \infty} \GG (x,r;x';r') \not\to 0 \ .
 \ee
A naive application of (\ref{uroP}) will yield a divergent answer
as the limit is not well-defined.

In order to obtain explicitly the  singular behavior of  $G(x,x')$ has to work with a
suitably regularized expression. To this end one can first consider some $x''$ lying in a small neighborhood of $x'$ in the boundary. If $x''$ is not connected to $x$ by null geodesics, (\ref{uroP}) can be straightforwardly applied to obtain $G(x,x'')$.  One can then take $x'' \to x'$
to obtain the desired correlator $G(x,x')$.

We will now motivate the above discussion in a slightly different
perspective using the geodesic approximation to $\GG (x,r;x',r')$.
For this purpose, we will take $m$ large\footnote{The relation
between null geodesics and singularities of $G(x,x')$ does not
require taking $m$ large.} so that $\Delta \approx m$. To consider
the limit (\ref{uroP}), we put the boundary $\bM$ at a cut-off
surface $r = r' = \Lam \to \infty$, on which (for an asymptotically AdS geometry)
the induced metric
can be approximated as
 \be \label{inm}
 ds^2   = \Lam^2 \, ds^2_{bd}
 \ee
where $ds_{bd}^2$ denotes the metric on the boundary $\bM$.

Consider points  $A =(x,\Lam)$ and $B = (x',\Lam)$ on the cut-off
surface, which are spacelike separated. In the large $m$ limit,
one expects that the saddle point approximation is valid. Denoting by
$d(A,B)$ the proper distance between two (spacelike separated)
points in $\CM$ we have
 \be \label{penB}
 \GG (A,B) \propto e^{- m \, d (A,B)} \sim e^{-2\, m \, \log \Lam} \ ,
 \ee
where we used the fact that as $\Lam \to \infty$, the proper
distance $d (A,B) \approx 2 \log \Lam + \cdots$ in asymptotically
AdS spacetimes.  Note that (\ref{penB}) is consistent with
(\ref{lim1}) and (\ref{lim2}) and hence makes sure that the limit
(\ref{uroP}) is well defined.\footnote{For generic timelike
separated points $A, B$ on $\bM$, we also expect (\ref{penB}) to
be true based on analytic continuation from spacelike separated
points. Note that in the pure AdS, $\GG$ is a function of the
proper distance $d (A,B)$ which can be analytically continued in
the complex plane to timelike separation.}

Now suppose that there is a null geodesic connecting $A$ and $B$ -- in
this case $d(A,B) = 0$ and $\GG(A,B) \sim O(1)$. As suggested
above we consider points $C = (x'',\Lam)$ in a small neighborhood
of $B$, for which
 \be \label{gej}
 d(A,C) \sim \log [\Lam^2 \,
 \delta x^2 \, ]
 \ee
where $\delta x$ is the proper distance between $x$ and $x''$
on the boundary (\ie, in terms of metric $ds_{bd}^2$ in (\ref{inm}) on $\bM$).
Plugging (\ref{gej}) into (\ref{penB}) and then taking the limit as suggested in
(\ref{uroP}), we have
 \be \label{sinB}
 G(x,x'') \sim {2m \ov (\delta x^2)^{m}}, \qquad x'' \to x' \ .
 \ee

To summarize, we have argued that a boundary correlation function
$G (x,x') = \vev{\OO(x)\OO(x')}$ is singular if and only if
\footnote{While we have explicitly shown that bulk null geodesics
lead to new singularities in boundary correlation functions, our
arguments indicate that the converse should also be true. This is
equivalent to the statement that (\ref{oep}) is true only for
points connected by null geodesics.} there exist null geodesics connecting $x$ and $x'$. Our argument is not completely rigorous and we discuss potential loopholes below. This proposal can be checked in simple examples in which one can
work out the boundary $G(x,x')$ explicitly, like global pure AdS,
Poincare patch and the BTZ black holes (see appendices).

\subsection{Types of null geodesics and causal structure}

In this subsection we discuss the location of the singularities in
$G(x,x')$ and structure of null geodesics. For this purpose, let
us fix $x $ and consider those $x'$ which are connected to $x$ by
null geodesics. The null geodesics can be divided into two types:
those lying entirely in the boundary (type \I) and those lying in
the bulk except for their end points (type \II). Type \I\ geodesics
are simply null geodesics of the boundary theory and define the
light cone of the CFT.  For points $x$ and $x'$ connected by Type \I\
null geodesics, $G (x,x')$ has the standard light-cone singularities.
Type \II\ null geodesics are more interesting and we now discuss them in some detail.

In pure global AdS (where the boundary is $\R \times \S^{d-1}$),
all Type \II\ null geodesics connect the anti-podal points on
$S^{d-1}$. The points in question are given by $(t_i, \Om)$ and
$(t_o, -\Om)$, with $t_o - t_i = \pi \, R_{AdS} = \pi$ and $\pm
\Om$ denote points which are anti-podal on the $\S^{d-1}$. Note
that the (boundary) time separation $t_o - t_i = \pi$ needed for
the geodesic to traverse the bulk is exactly the same as that of a
Type \I\ null geodesic connecting the anti-podal points only in
pure AdS spacetime; the former becomes longer for general
asymptotically AdS spacetimes. Further details on geodesics in
AdS, both analytic and pictorial, appear in \App{appAdS}.

Since in the case of pure AdS the end points of the Type \II\ null
geodesics already coincide with the boundary light cone, they do
not give rise to additional singularities in the boundary
correlation functions.  It is easy to verify that the behaviour of
null geodesics in AdS is consistent with the singularity
properties of boundary correlation functions. The boundary
correlators in questions are two-point functions in the CFT vacuum
of primary operators of dimension $\Delta$ which for a
$d$-dimensional CFT on $\R \times \S^{d-1}$ are given by
\eqn{adscorr}{
\vev {\CO\(t,\Om\) \, \CO\(t',\Om'\) } \propto {1 \over
\(\cos(t-t') - \cos(\Om -\Om') \)^{\Delta} } \ , }
The singularities of the correlation function \adscorr\ are given
by $\delta t = \delta \Om$ with $\delta t = t-t'$, \etc, which
coincide with the light cone of the boundary manifold.

In a perturbed AdS spacetime, the end points of Type \II\
geodesics in general do not coincide with the light cone of the
boundary manifold. They will therefore give rise to additional
singularities in the boundary correlation functions. We will call
such singularities {\it bulk-cone singularities.} An interesting
feature of asymptotically AdS spacetimes is that there exists a
lower bound on the `time delay' experienced by geodesics exploring
the bulk spacetime. In a situation where the bulk manifold $\CM$
with a timelike conformal boundary $\bM$ satisfies the null energy
condition, the null generic condition, and strong causality and
compactness, it has been proven by Gao and Wald~\cite{Gao:2000ga}
that the end points $x'$ of Type \II\ geodesics always lie {\it
inside} the light cone of the boundary manifold, \ie, $x'$ and $x$
are timelike separated. This implies that any generic perturbation
of AdS (satisfying the conditions mentioned) will produce a time
delay of null geodesics relative to pure AdS. For example, imagine
a state corresponding to a thermal gas in AdS at some internal
mass density $\rho_o$. Then a radial geodesic will arrive to the
anti-podal point of the boundary after a time $\Delta t$ which
increases monotonically with increasing $\rho_o$.

This implies that correlation functions in some excited
states\footnote{In order to have non-negligible back reaction on
the bulk metric, the ADM mass $M$ of the perturbation should
satisfies $G_N M \sim \CO(1)$. This implies that the corresponding
excited state in boundary theory should have energy of order
$\CO(N^2)$. In the following discussion, by generic excited states, we mean generic
states of such energies.} of the theory will have additional
Lorentzian singularities inside the light cone.
We also note that it can be checked explicitly that for
backgrounds with non-compact boundary, Type \II\ null geodesics do not exist for states preserving the boundary symmetries (for a proof of the statement see \App{apppoincare}). Thus additional singularities appear to arise only for CFTs on compact spaces or those that violate the Poincar\'e symmetery for CFT on $\R^{3,1}$.

While the theorem of Gao and Wald only applies to asymptotically AdS
spacetime satisfying the conditions stated above, from the
causality of the boundary theory one would like to conclude that
the end points of Type \II\ null geodesics should always lie inside
the light cone of the boundary for any asymptotically AdS background
with a field theory dual. Otherwise it would imply that one can set up
a special state in the field theory to transmit signals faster
than the speed of light (see also~\cite{Kleban:2001nh}).

To conclude this section let us comment on a potential loophole in
our argument. Our connection between null geodesics and
singularities of boundary correlation functions is based on the
validity of (\ref{oep}) for any two points connected by a null
geodesic. While this assumption is consistent with our general
understanding of quantum field theory in a curved spacetime, we do
not have a rigorous proof of the statement.\footnote{For example,
in a semi-classical approximation to the propagator, one may
consider the following scenario: In general, the end points of a
null geodesics can be shared by other spacelike and/or timelike
geodesics. If the propagator is given by summing over all
geodesics, then the null geodesic clearly dominates. However,
there might be a situation that path integration contours for the
propagator cannot be deformed to the saddle point corresponding to
the null geodesic. Then the null geodesic will not contribute to
the propagator, even though it naively dominates.
Analogous phenomenon was encountered in \cite{Fidkowski:2003nf} for a spacelike (almost null) geodesic `bouncing off' a black hole singularity; though
it is not clear whether this contingency will ever arise for strictly null geodesics.
Given that
one can send light signals following null geodesics, one might be
able to argue a null geodesic should always contribute to a
propagator.} Also note that one should distinguish between points
which can be connected by a null geodesic and those points which
the geodesics connecting them only {\it approach} null. In the latter
case, when there does {\it not} exist a null geodesic which connects
two points, our argument does not directly apply. We will
elaborate more on this when discussing the gravitational collapse
background in \sec{horform}.

\section{Static asymptotically AdS geometries }
\label{statgeom}

We now turn to applying the results of \sec{nullgsing} to static
spacetimes. A trivial example where our relations can be verified
is of course the empty AdS spacetime which we have already commented on.
We now turn to consider geometries which are deformations of AdS, with a free parameter such that pure AdS is retrieved in a particular limit of this parameter.
We will first
consider situations with sufficient symmetries to simplify the analysis before
turning to more complicated examples.
For concreteness, we will work in 5 dimensions, \ie, take the spacetime to be asymptotically $AdS_5$ (times a $\S^5$ which will not play any role).

\subsection{Geometries of stars in AdS}

As our first nontrivial example, we will take our spacetime to be
static and spherically symmetric, and to make things even simpler,
we will take the matter content to be that of a self-gravitating
gas of radiation. We note that self-gravitating radiation in AdS has been explored previously in \cite{Page:1985em} and related geometries of boson stars in asymptotically AdS spacetimes have been considered in \eg\ \cite{Astefanesei:2003qy,Astefanesei:2003rw}. The symmetries imply the corresponding stress
tensor is that of a perfect fluid; radiation equation of state
which makes the stress tensor traceless, \ie,
 \eqn{Tstar}{
T_{ab} = \rho (r) \, u_a \, u_b + P (r) \, \( g_{ab} +  u_a \, u_b
\), \qquad  P = {1 \over 4} \, \rho  \ }
where $u^a$ is a co-moving gas 4-velocity, and $\rho(r)$, $P(r)$ are density and pressure. Since the radiation will be confined by the AdS potential, we will refer to this configuration as a ``star" in AdS.  The metric can be obtained by solving the Einstein's equations with negative comological constant and the requisite stress tensor, as discussed in detail in \App{appstar}. Here we summarize the main results.

The metric can be written in the form:
\eqn{starmet}{ ds^2 = -f(r) \, dt^2 +  h(r) \, {dr^2} + r^2 \,
d\Om_{3}^2 \  }
with
 \eqn{hstar}{ h(r) = \[ {r^2 } + 1 - {m(r) \over r^2}
 \]^{-1} }
 and
 \eqn{afstar}{ f(r) = \( { \rho_{\infty} \over
\rho(r) } \)^{\! 2/5} }
 where the mass function $m(r)$ is defined in
terms of an integral of the density $\rho(r)$,
 \eqn{mstar}{ m(r) \equiv {2
 \over 3} \, \int_0^r \rho(\rb) \, \rb^3 \, d\rb
  }
and $\rho_{\infty}$ is the coefficient of the leading fall-off of
$\rho(r)$, determined by $\rho(r) \sim {\rho_{\infty} / r^5}$
as $r \to \infty$.  The field equations specify the system
of coupled first order ODEs which  $m(r)$ and $\rho(r)$ must satisfy:
\begin{eqnarray} \label{m1}
&  m'(r) = {2 \over 3} \, \rho(r) \, r^3 \\
 \label{p1}
& \rho'(r) = -5 \, {\rho(r) \over r} \,  \( {
 {r^2 } + {m(r) \over r^2} +{1 \over 12} \, \rho(r) \, r^2 \over
 {r^2 } + 1 - { m(r) \over r^2} } \)
\end{eqnarray}
with boundary conditions $m(0)=0$ and $\rho(0) \equiv \ro$, where
$\ro$ is a free parameter of the configuration, specifying the
internal density of the gas. Thus the geometries are parameterized
by a single parameter $\ro$, with pure AdS retrieved in the $\ro = 0$ case.

\begin{figure}[htbp]
\begin{center}
\includegraphics[width=6in]{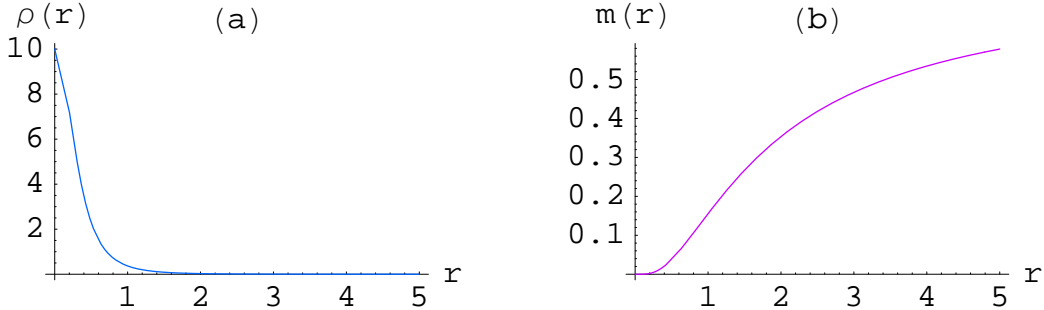}
\caption{Density and mass functions for the ``star" geometry,
where the central density is set to $\ro=10$.} \label{starrhom1}
\end{center}
\end{figure}

From (\ref{m1}) and (\ref{p1}) we find the following asymptotic
behavior for $m(r)$ and $\rho (r)$
 \bea
  r \to \infty: \qquad && \rho(r) \to
 {\rho_\infty \ov r^5} \(1- {5 \ov 2r^2} +  {5 (4M +7) \ov 8 r^4} + \cdots \) , \\
  &&  m (r) \to M - {\rho_\infty \ov r} + \cdots \\
  r \to 0: \qquad && \rho(r) \to \rho_0 + O(r^2), \qquad
 m (r) \to {\rho_0 \ov 6} r^4 + O(r^6)
 \eea
where $\rho_\infty$ and $M$ ($M$ is proportional to ADM mass) are
constants which are determined by $\ro$. Using (\ref{hstar})
and \afstar\ we also find that
 \bea
 && r \to \infty: \qquad f(r) \to r^2 + 1 - {M \ov r^2} + \cdots \qquad
 h (r) \to \[ {r^2 } + 1 - {M \over r^2} + \cdots
 \]^{-1}   \\
 && r \to 0: \qquad f(r) \to \({\rho_\infty \ov \rho_0}\)^{2 \ov 5} + O(r^2), \qquad
 h (r) \to 1 + O(r^2)
 \eea

One can readily solve the full equations (\ref{m1}) and (\ref{p1})
numerically.  \fig{starrhom1} shows the density and mass
functions for the internal density $\ro=10$.

\begin{figure}[htbp]
\begin{center}
\includegraphics[width=6in]{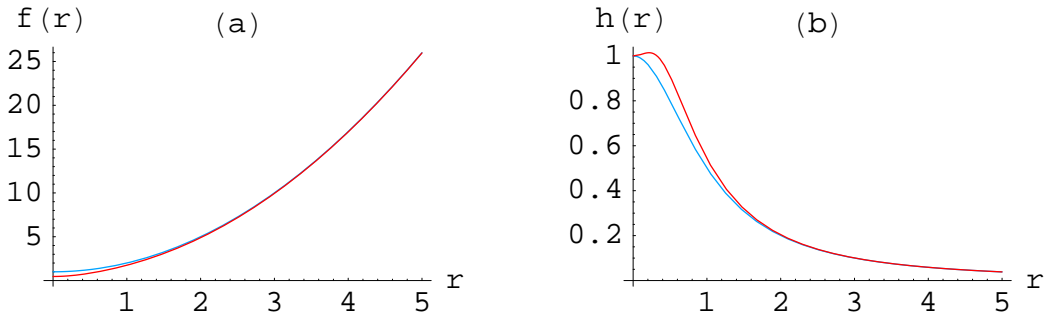}
\caption{Metric functions for the ``star" geometry with central
density $\ro=10$ (red curves), and for comparison corresponding
metric functions in the pure AdS geometry (blue curves; $f(r)$ is
the higher and $h(r)$ is the lower of the two curves).}
\label{starfh1}
\end{center}
\end{figure}
The metric functions are plotted in \fig{starfh1}, again
for the same set-up as in \fig{starrhom1}.  For comparison we
also plot the corresponding metric coefficients for pure AdS; as
we would expect, the metric does approach that of AdS as the
density becomes small.

\subsection{Geodesics in the star AdS geometry}

We will now turn to the properties of bulk null geodesics (Type
\II) in the star geometry. Staticity and spherical symmetry implies
that $E$ (energy) and $J$ (angular momentum) are conserved along
any geodesic and hence we will parameterize the geodesics by these
quantities. The equations for null geodesics are:
\eqn{geodeqs}{ \dot{t} = {\al \over f(r)} \ , \qquad \dot{\ph} =
{1 \over r^2} \ , \qquad {\rm and} \qquad f(r) \,  h(r) \,  \dot{r}^2 = \al^2 -
V(r)  \ , }
where
\be V(r) =  {f(r) \over r^2}  , \qquad \al \equiv {E \ov J} \ee
and the dot denotes differentiation with respect to the affine
parameter along the geodesic. For null geodesics the absence
of scale implies that the relevant parameter is the ratio $\al =
E/J$. Useful quantifiers for figuring out which points on the
boundary are connected by null geodesics are the temporal and
angular separation of the geodesic endpoints, which we denote by
$\D t$ and $\D \ph$, respectively. Since we have a numerical
solution for the metric, we have to integrate the geodesic
equations numerically to find the desired properties.

Consider first radial null geodesics, which we expect will emerge
at a later time then the corresponding geodesic in pure AdS,
consistent with the theorem of \cite{Gao:2000ga}.  Denoting this
time by $\Delta t_0$, we have \eqn{starDt}{ \Delta t_0 = 2 \,
\int_0^{\infty} \sqrt{{h(r) \over f(r)}} \, dr }
\begin{figure}[htbp]
\begin{center}
\includegraphics[width=4in]{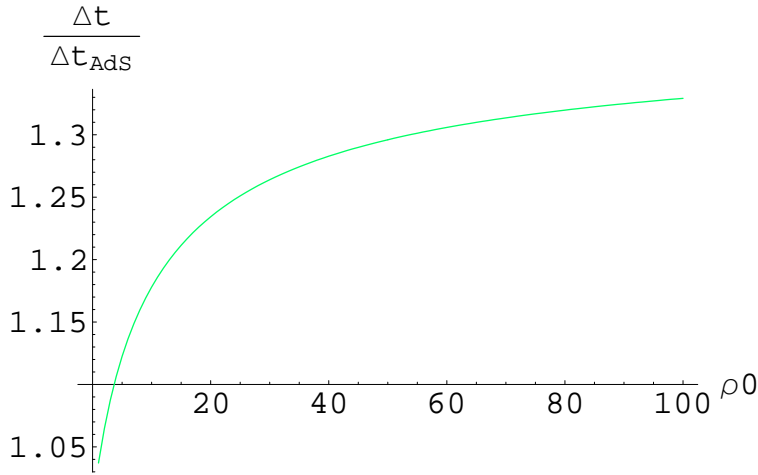}
\caption{Time delay for radial null geodesics through star in AdS, as a function of the star's internal density $\ro$. }
\label{startdelay}
\end{center}
\end{figure}
For  a star geometry with central density $\ro=10$, we find
numerically that $\Delta t_0 \approx 3.696$, which is 1.177 times
longer than in pure AdS.  \fig{startdelay} exhibits the dependence
of this time delay on the value of $\ro$.  We see that while the
time delay increases monotonically with increasing density, it
does so more and more slowly.\footnote{The intuitive reason for
$\Delta t_0$ not increasing more rapidly with $\ro$ is that the
larger the internal density $\ro$, the faster the density function
$\rho(r)$ falls off, in such a way as to keep the total mass $M$
bounded, as discussed in \App{appstar}. Numerically we find that
$\Delta t_0$ appears to be bounded as $\rho_0 \to \infty$.}

\begin{figure}[htbp]
\begin{center}
\includegraphics[width=5in]{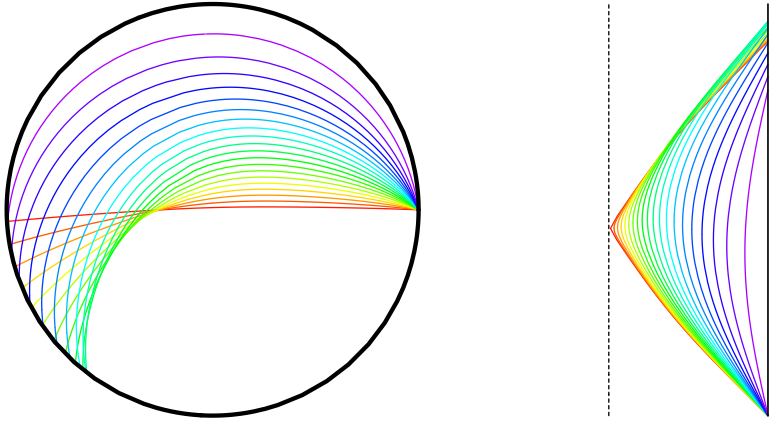}
\caption{Null geodesics in star with $\ro=10$ in AdS, projected
onto  a constant $t$ slice and the $t-r$ plane, for varying
angular momentum to energy ratio  ($E = 10$ and $J= 0,1,\ldots,10$).
On the left, the bold circle corresponds to the AdS boundary, whereas on the right, the dashed vertical line corresponds to the origin $r=0$ and the
bold vertical line to the boundary at $\tan r = {\pi \over 2}$.
The range of $t$ plotted is $(0, 1.1 \, \Delta t_0)$.  (Analogous plots with larger $\ro$ appear in \fig{starrphrho} of \App{appstar}.)}
\label{stargeodsnull}
\end{center}
\end{figure}

Now let us consider null geodesics with nonzero angular momentum.
(\App{appstar} further explores non-radial spacelike geodesics.)
In \fig{stargeodsnull}, we show null geodesics projected
onto a spatial (constant $t$) slice (left) and the $t-r$ plane (right).
Not surprisingly,
the star has a focusing effect on the geodesics.
 The important point to note is that now, unlike the pure AdS
case, there is a finite spread of the $\D t$ and $\D \ph$ endpoint
values the null geodesics can take.  More specifically, because of the
attractive nature of the star's gravitational potential, the small
angular momentum ($J/E \ll 1$) geodesics `overshoot' in terms of
angular separation, and at the same time they experience some time
delay, which means that both $\D t$ and $\D \ph$ increase.
  However, this effect does not keep increasing
monotonically with increasing angular momentum, because the
further from the origin the geodesics penetrate, the less effect of the
star they feel.  In the high angular momentum limit ($J/E \sim
1$), the geodesics hug the boundary and consequently behave just
as the corresponding geodesics in pure AdS.

\begin{figure}[htbp]
\begin{center}
\includegraphics[width=6in]{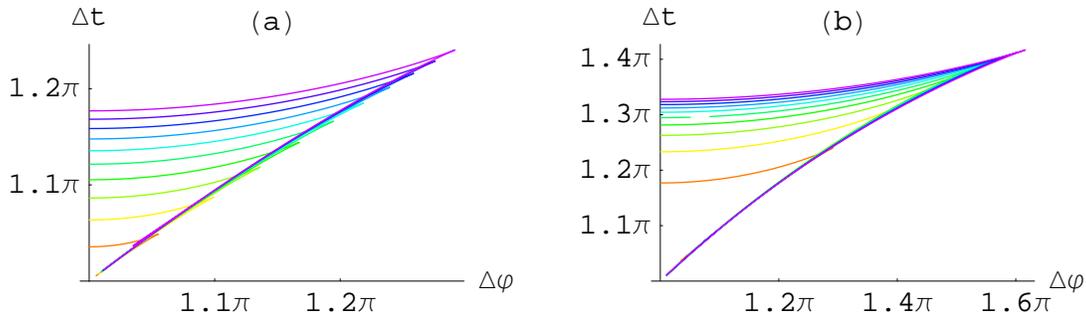}
\caption{Endpoints of null geodesics in the AdS geometry of star
with {\bf (a)} $\ro=1, 2,\ldots,10$ and  {\bf (b)} $\ro=10, 20,\ldots,100$.
Each curve is plotted by varying $\al={E \ov J}$.
The top of the curves corresponds to big values of $\al$;
in fact, one can easily show that the slope of each curve
(as a function of $\al$) is given by $1/\al$.}
\label{startphrho}
\end{center}
\end{figure}

To illustrate explicitly how the
endpoints of these various geodesics compare,
\fig{startphrho} summarizes the behaviour of $\Delta t$ and $\Delta \ph$
for the various geodesics.
In (a), where the endpoints are plotted for small internal densities, $\ro = 1, \ldots, 10$, we see that the effect of overshooting becomes more pronounced the higher $\ro$ is.  In (b), where $\ro = 10, \ldots, 100$,  we illustrate that the effect however remains bounded even for high internal densities.  (We have verified this numerically for $\ro \lesssim 10^{18}$.)  Note that the $\D t$ intercept for each curve can be read off already from \fig{startdelay}.

The boundedness of $\D t$ implies one of the interesting features for the star geometries, namely the absence of null geodesics orbiting the star.   In fact, one can obtain this result more directly, as follows:
From \geodeqs\ we see that in order to have null geodesics at fixed radial distance from the star,
 \ie, $r(\lambda) = {\rm constant}$,
we require that $\a^2 - V(r) =0$ and ${d \over dr } V(r)  =0$.
It is easy to show numerically that these conditions are never satisfied for the star geometries.\footnote{The two conditions can be distilled into a necessary condition for circular photon orbits;
defining $Q(r) = {1 \over 12} \, \rho(r) \, r^2 + 2 \, {m(r) \over r^2}$
we need  $Q(r_0) = 1$ at some $r =r_0$. We find that $Q(r) \lesssim 0.6$ for all $r$ irrespective of the internal density $\rho_0$.  }
As we will argue below, this absence of circular orbits around the star leads to marked differences in the behaviour of singularities for correlation functions in the pure state corresponding to the star and in themal
density matrix represented by an eternal black hole geometry.
In particular, in the star case, the spread of the geodesic endpoints remains bounded, unlike for the black hole case which has a circular orbit.

The essential points relating to null geodesics in the star background can be summarized as follows:
\begin{itemize}
\item The radial null geodesics which go through the bulk geometry emerge at the anti-podal
point on the sphere ($\D \ph = \pi$) at a later time than in pure AdS ($\D t > \pi$). The time delay is a monotonic function of
the internal density $\rho_0$.
\item Null geodesics with angular momentum exhibit both a time delay as well as a shift in $\Delta \ph$.  For small angular momentum, both $\Delta t$ and $\Delta \ph$ increase with increasing angular momentum, whereas for large angular momentum they decrease to match with the AdS value $\Delta t = \Delta \ph = \pi$ when $J=E$.  While this effect increases with increasing $\ro$, $\Delta t$ and $\Delta \ph$ remain bounded for arbitrary $\ro$.
\item There are no null geodesics at fixed value of the radial coordinate \ie, the star spacetime does not admit circular photon orbits.
\end{itemize}

We stress that since the endpoints of the null geodesics indicate the location of a singularity in the corresponding correlation function as discussed above, the data of \fig{startphrho} is in principle easily extractible from the gauge theory.  This demonstrates how from the gauge theory we can read off the details of the geometry deep inside the bulk.  In other words, just by looking at the spread of the endpoints of the null geodesics, we can distinguish the star from a small black hole (as demonstrated below) or other configurations of the same mass in AdS.  Note that this is rather different method of extraction of details of deep IR in the bulk from that employed in \cite{Horowitz:2000fm}: the latter considered one-point functions, whereas here we make explicit use of the nonlocal nature of the two-point functions.

\subsection{Field theory signature of excited AdS geometries}

We now turn to a brief analysis of the star geometry from a field
theory perspective. By construction we have an asymptotically AdS
geometry which is static and spherically symmetric. While we have
modeled the star by an effective equation of state, it is easy to
see that similar solutions can be obtained in gauged supergravity
theories. In any event the geometries in question correspond to
normalizable deformations of the AdS vacuum, implying that they
are dual to states in the field theory. In particular, these
states should be eigenstates of the CFT Hamiltonian (since they
are static). One can therefore use the state operator
correspondence  in the CFT  to map the state into an operator
insertion; for convenience we will call the state of the star
$\ket{star}$ and the corresponding operator $\CO_{*}$ \ie,
$\ket{star} = \CO_* \, \ket{0}$. Note that to obtain a geometry
with spherical symmetry we would need to smear the operator
insertion $\CO_*$ on the boundary $\S^3$ making it effectively
non-local.

One of the interesting questions that we will be unable to provide
a concrete answer for is the precise description of the state
$\ket{star}$ in the field theory. From general arguments it is
clear that the operator $\CO_*$ has dimension of $\CO(N^2)$ in the
field theory.\footnote{ This is because of the state being dual to
a configuration of finite ADM mass $M$ and the fact that the bulk
Newton's constant scales like $N^2$.}  In order to ascertain the
precise state we need more information than the one point function
of the stress tensor (which is of course given by the leading
fall-off in the metric). However, it is clear that on physical
grounds such states ought to be constructible in the field theory
and we will proceed with the assumption that $\CO_*$ is a {\it
bona fide} operator in the field theory.

Our analysis of geodesics in the star spacetime can be rephrased
in light of the discussion in \sec{nullgsing} as predictions for
new `light-cone' singularities in the boundary correlators in the
state $\ket{star}$. The  later arrival at the boundary of radial
null geodesics consistent indicates that correlation functions of
generic operators $\CO$ in the field theory, $\bra{star} \, \CO(x)
\, \CO(y) \, \ket{star}$  will have singularities at $\Delta \ph
= \pi$ and $\Delta t $ given by the appropriate time delay.
Similarly, the implication of bulk Type \II\ geodesics with angular
momentum is that we will have light-cone singularities in field
theory correlators at operator insertion points $x, y$ such that
$t_x - t_y = \Delta t$ and $\Om_x - \Om_y = \Delta \ph$ with
$\Delta t$ and $\Delta \ph$ given by the geodesic analysis.

While we have emphasized the results here as predictions for the
gauge theory correlation functions in an excited state at strong
coupling, we believe that some of these predictions can be tested.
For instance, one can consider the $1/2$ BPS states in $\CN =4$
SYM. The geometries dual to these states are explicitly
constructed in \cite{Lin:2004nb}. Within this class of geometries
one can focus on those which are asymptotically AdS and calculate
the bulk prediction for the light-cone singularities using the
nature of bulk geodesics. To compare with the field theory result
we need to know the behavior of four point functions, which can be
analyzed in large $N$ perturbation theory (at weak coupling) by
taking the probe operators to also be chiral primary. We hope to
report on this in the near future.

\section{Eternal black holes in AdS}
\label{bhthermal}
In the previous section we discussed the behaviour of geodesics in spacetimes which are deformations of AdS, focussing on globally static geometries. We will now consider spacetimes such as the eternal \SAdS\ black hole which is causally nontrivial and not globally static. In this case we know that a large \SAdS\ black hole corresponds to a thermal state in the field theory. The analysis of geodesics will lead to interesting signatures of the black hole geometry which will be visible in thermal correlation functions.

\subsection{\SAdS\ black hole: Structure of spacelike and null geodesics}
We will first describe the structure of null and spacelike
geodesics in an AdS black hole background. Again focusing attention on 5-dimentional AdS, the black hole metric
is given by
 \be
 ds^2 = -f(r) \,  dt^2 + {1 \ov f(r)} dr^2 + r^2 \, d \Om_{3}^2
 \ee
 with
 \be \label{schwfn}
 f(r) = r^2 +1 - {\mu \ov r^2} \ .
 \ee
The mass of the black hole is proportional to $\mu$, and correspondingly the horizon
radius $\rh$ and the inverse Hawking temperature $\beta$ are given
by\footnote{Note that to describe a big black hole we need $\mu
\geq {3 \ov 4}$. Further, for the temperature of the system to
be high enough so that the big black hole dominates the thermal
ensemble, we require $\mu \geq 2$, which corresponds to $T = {1 \ov
\beta} \geq {3 \ov 2 \pi}$. For illustration, we often use
an example with $\mu=1$ below.}
 \be
 \rh^2 = {\sqrt{1+ 4 \mu} -1 \ov 2}, \qquad \beta = {2 \pi\, \rh \ov
 2\, \rh^2 +1} \ .
 \ee

The equations for general geodesics are given as:
\eqn{geodbhs}{ \dot{t} = {\a \over f(r)} \ , \qquad \dot{\ph} = {1
\over r^2} \ , \qquad {\rm and} \ \ \dot{r}^2 = \a^2 - V_{eff}(r)  \ ,
}
where
\eqn{bheffv}{
V_{eff}(r) = - {\k \over J^2} \, f(r) + \, {f(r) \over r^2}  \\
}
and $\k = 1,0,$ or $-1$ for spacelike, null, or timelike
geodesics, respectively, and the dot denotes differentiation with
respect to the affine parameter along the geodesic.

\begin{figure}[htbp]
\begin{center}
\includegraphics[scale=0.8]{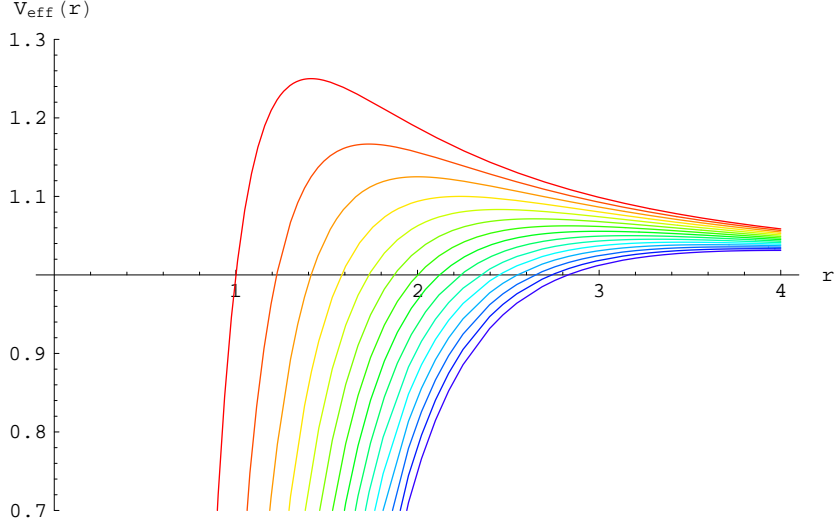}
 \caption{\small {The plot for $V_{eff}(r) = f(r)/r^2$ for $\mu \in \{1,1.5,2,\cdots,7.5,8\}$.}}
 \label{vpot1}
 \end{center}
\end{figure}

\begin{figure}[htbp]
\begin{center}
\includegraphics[scale=1]{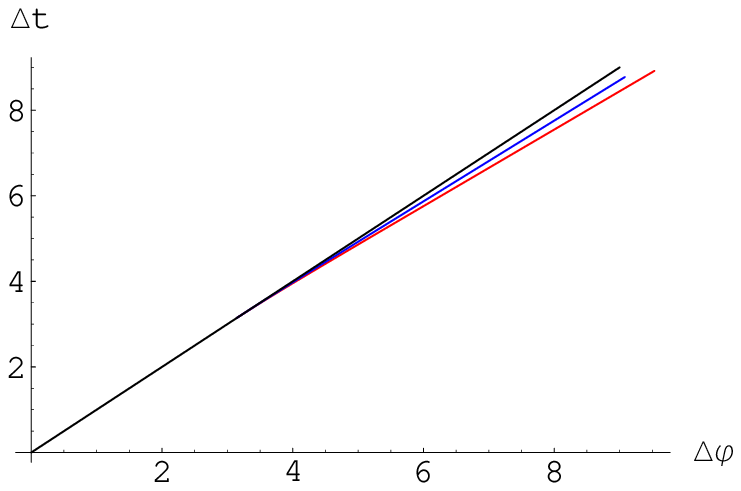}
 \caption{\small {End points of null geodesics in a black hole
 background. The black line is $\Delta t = \Delta \ph$, which is the
set of end points of boundary null geodesics. The blue and red
curves give the end points of bulk null geodesics for a black hole
with $\mu=2$ and $\mu=1$, respectively. It is obtained by varying
$\al \in (1,\am)$. As $\al \to 1$, $(\Delta t, \Delta \ph) \to
(\pi,\pi)$ and  as $\al \to \am$, $ \Delta t$ and $\Delta \ph$
approach infinity with their ratio given by $1/\am$. Note that
$\Delta \ph (\al)
> \Delta t(\al)$ for $\al \in (1,\am)$. }}
 \label{yp}
\end{center}
\end{figure}

Null geodesics in the black hole background ($\k =0$) are
parameterized by the ratio $\a = {E\over J}$.  Once again we will
be interested to know which points are connected by bulk null
geodesics and we will quantify this in terms of  $\(\D t (\al), \D
\ph (\al)\)$. The radial equation is easy to  interpret as a
classical particle with energy $\al^2$ moving in a potential
$V_{eff}(r) = {f(r) \ov r^2}$ (see \fig{vpot1}).
$V_{eff}(r)$ has a maximum at $r^2_m = 2\mu$ with a value $V_{eff}(r_m) =
1 + {1 \ov 4 \mu}$ and as $r \to \infty$, $V_{eff} \to 1$. The
null geodesics connecting boundary points are:

\begin{enumerate}

\item Geodesics with $\al =1$, \ie, $E = J$, stay at constant
$r=\infty$; these are null geodesics of the boundary manifold (Type \I).

\item The null geodesics which pass into the bulk and come back to
the boundary (Type \II) exist only for  $\al \in (1,
\am)$, where
 \be \label{deal}
 \am^2 = V_{eff} (r_m) = 1 + {1 \ov 4 \mu}
 \ee
As $\al \to 1$, we find that  $(\D t, \D \ph)  \to (\pi,\pi)$. It is also clear from \fig{vpot1} that as $\al \to \am$, $ \D t (\al)$ and $ \D \ph (\al)$ should go to infinity since it takes infinite affine parameter time to reach the turning point.\footnote{Or said differently, as $\al \to \am$, the geodesic can go around the circular orbit at $r_m=\sqrt{2 \mu }$ many times before re-escaping to the boundary.} One can further check
that
 \be \label{rhe}
\al \, {d t \ov d
\al} = {d \ph \ov d \al} \qquad \thus \qquad
 {d \ph \ov d t} = \al (t) .
 \ee
From this we conclude that as  $\al \to \am$,
 \be \label{rjrr}
  \D t (\al) \approx {1 \ov \am} \, \D \ph (\al) \to \infty
 \ee
Note that since $\am > 1$, the above equation implies $\D t
< \D  \ph$. This remains true for general values of $\al \in
(1, \am)$ as is clear from \fig{yp}, which is a parametric
plot of $\D \ph(\al)-\D t (\al)$ by varying $\al$. Note that since
$\ph$ is periodic, there is no violation of causality here.

\end{enumerate}

In order to better understand the set of boundary points that can
be connected through the bulk, we should also look at spacelike
geodesics, which are discussed in detail in
Appendix~\ref{BHspace}.

\subsection{New singularities for thermal YM theories on $\S^3$}

We now consider gauge theory implications of the null geodesics in the black
hole background. As discussed in \sec{nullgsing}, we will assume that null geodesics lead to the singularities of boundary correlation functions.

At a spatial point $\D \ph= \ph_0$, boundary null geodesics will
give rise to singularities at
 \be \label{oldsin}
 \D t = \ph_0, \; \ph_0 + 2 \pi, \; \ph_0 + 4 \pi, \cdots \ .
 \ee
These are singularities of vacuum correlation functions and we expect they
survive at finite temperature as well. Type \II\  null geodesics which go
into the bulk will give rise to new singularities at
 \be \label{newsin}
 \D t_{1} =\ph_0 + {2 \pi  \; b_1 (\ph_0)}, \; \D t_{ 2} = \D t_1 + {2 \pi \; b_2
 (\ph_0)}, \; \cdots, \; \D t_{  n} = \D t_{  n-1} + {2  \pi  \; b_n  (\ph_0)}, \;
 \cdots,
 \ee
where $\D t_n,\;  n=1,2,\cdots$ are obtained by the intersection of
the red line with vertical lines $\D \ph = 2 \pi n + \ph_0$ in
\fig{yp}. Here $b_n (\ph_0)$ is some number lying between
$1/\am$ and $1$ and is smaller than $1$. As $n$ becomes large,
 \be
 b_n (\ph_0) \to {1 \ov \am} = {1 \ov \sqrt{ 1 + {1 \ov 4\;  \mu}}}
 \ee
independent of the value of $\ph_0$.
The locations of these singularities are temperature-dependent, since  $\am$
and  $b_n$ depend on the temperature of the system (through $\mu$).
Since all $b_n$ are smaller than 1, the singularities in (\ref{newsin}) are generically
distinct\footnote{At some special values of $\ph$ and $t$, it is
possible for them to coincide.}  from those in (\ref{oldsin}).
Note as we increase the temperature of the system (\ie, increase
the black hole mass $\mu$), $\am \to 1$ and the two sets of
singularities merge in the $\mu \to \infty$ limit. In the high
temperature limit, $\S^3$ effectively decompactifies to $\R^3$, and
as we discussed earlier there are no bulk null geodesics in this
limit (see \App{apppoincare}) when we insist on maintaining the Poincar\'e symmetry.

Since ``new'' singularities arise from null geodesics which go
into the bulk, they encode information regarding the bulk
geometry. That singularities exist for $\D t \to \infty$ is a
reflection of the existence of a (unstable) circular orbit at
$r_m =\sqrt{2 \, \mu}$ in the bulk geometry. The time difference $\D
t_n-\D t_{n-1}$  between nearby singularities for $n$ large is
simply given by the time that it takes a
null geodesic to transverse the circular orbit (which is $2 \pi/\am$). More generally,
denoting the locations of the singularities by the curve $\D t(\D \ph)$,
 from the second equation of (\ref{rhe}) it follows that the slope ${d \D t \ov
d \D \ph}$ gives $1/\al =J/E$ of the corresponding bulk geodesic.

Also note that in cases where the bulk geometry deviates from that
of the \SAdS\ geometry in the vicinity of the horizon, such as the
smooth microstate geometries discussed in the context of D1-D5 systems (\cf, \cite{Mathur:2005zp, Mathur:2005ai} for reviews and references), the
gauge theory should in principle be able to detect even slight
deviations which cause a slight shift in the radius of the null
circular orbit. This is facilitated by the fact that the slope ${d
\D t \ov d \D \ph}$ can be measured arbitrarily precisely for
large enough separations $\D t$ and $\D \ph$.

Our conclusion above was obtained in the supergravity limit, \ie,
large $N$ and large $\lam$ limit. It would be interesting to
understand whether the ``new singularities'' at finite temperature
arise when one departs away from the limit. In particular, it
would be interesting if one could find independent arguments for
their existence in the gauge theory.

\section{Horizon formation from gauge theories}
\label{horform}

In the preceding sections we have focused on static\footnote{Although the eternal black hole spacetime is not globally static, the method sketched above only probed the static part of the geometry.  We comment on extracting the details of the dynamic spacetime inside the horizon in \App{BHspace}.}
bulk geometries and indicated how to extract certain details about the geometry from the boundary correlators. In this section we finally apply the relation between null geodesics and singularities of boundary correlators to the fully dynamical case of gravitational collapse. Previous studies of black hole collapse in the AdS/CFT context are \cite{Danielsson:1999zt,Giddings:1999zu,Giddings:2001ii,Maeda:2006cy}.

The basic question we want to ask is: {\it Can we see horizon
formation directly in the gauge theory?} Since in the bulk,
horizon formation is a sharply-localized event (despite the fact
that we need to know the full future evolution of the spacetime to
actually find the horizon), we would expect that there will
correspondingly be some sharp feature in the gauge theory. We will
start by considering the simplest toy model, which is that of a collapsing
spherical null shell. However, our main result should be
applicable to general gravitational collapse, as will be clear
below.

\subsection{Null geodesics in a gravitational collapse}

 Consider a null spherical shell in $d$-dimensional asymptotically AdS spacetime.
We can take the metric inside and in the past of the shell to be pure
AdS, and the metric outside and to the future of the shell to be
\SAdS.  The metrics for the interior and the exterior of the shell
can be written as
\eqn{metricgen}{ ds^2 = -f_{in,out}(r) \,
dt_{in,out}^2 + {dr^2 \over f_{in,out} (r)} + r^2 \, d\Om_{d-2}^2}
where
\begin{eqnarray}
\label{fin}
 &f_{in} (r) = {r^2} + 1
\\
\label{fout}
 &f_{out} (r) =  {r^2 } + 1 - {\mu \over r^{d-3}}
\ .
\end{eqnarray}

The mass of the black hole is proportional to $\mu$. As in the previous section, we will use
$r_+$ to denote the horizon radius and the surface gravity
$\kappa$ of the black hole is given by
 \be
 \kappa = {2 \pi \ov \beta} = \ha {d f_{out} \ov dr} \biggr|_{r_+}
 \ee
with $\beta$ the inverse Hawking temperature. Note that the $t$
coordinate jumps across the shell with the jump approaching zero
at the boundary. In contrast, the $r$ coordinate is physical
(since it measures the proper area of the spheres in a spherically
symmetric spacetime) and therefore varies continuously across the
shell.

\begin{figure}[htbp]
\begin{center}
\includegraphics[width=6in]{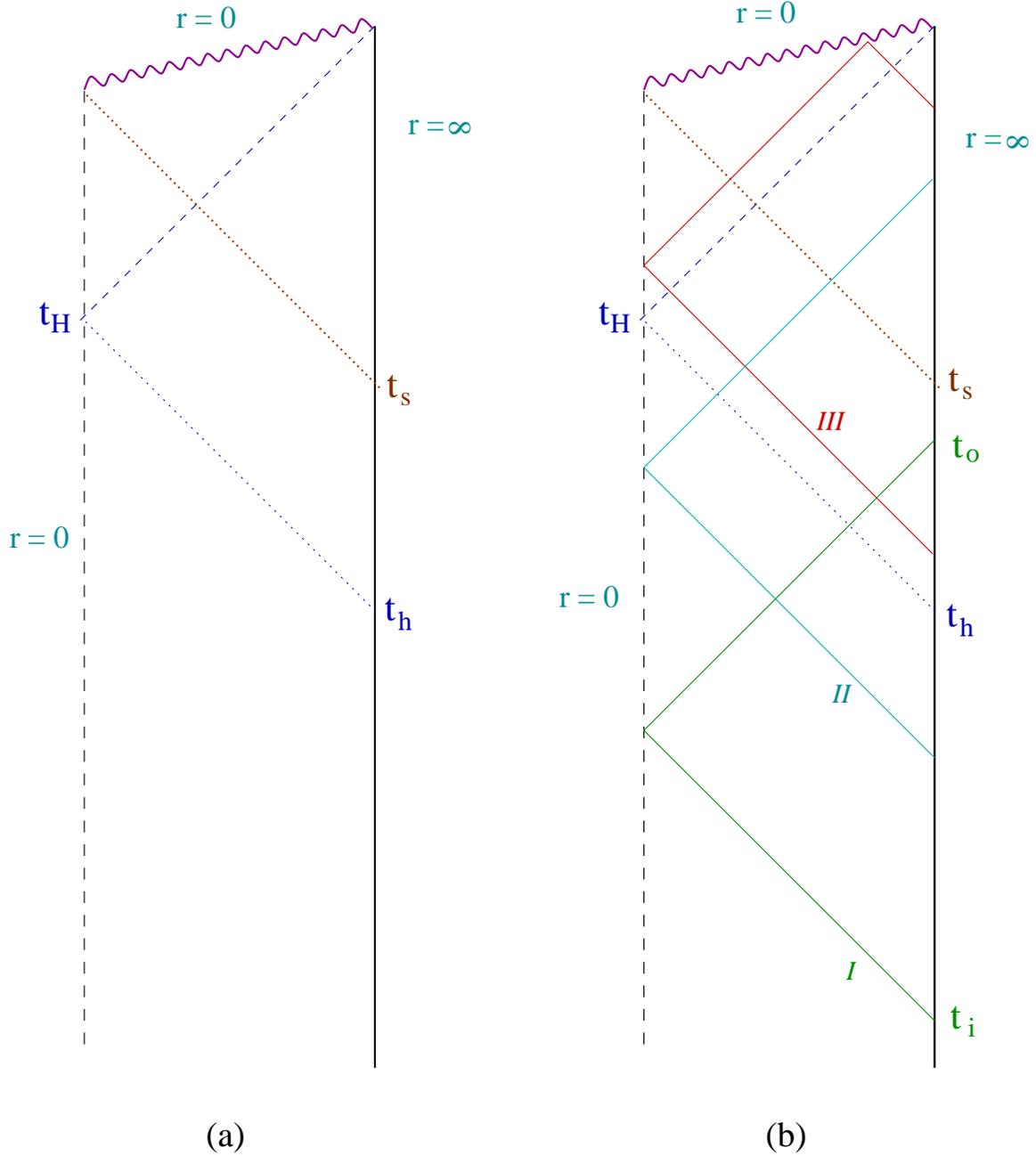}
\caption{Penrose diagram for a collapsed black hole.  {\bf (a)} shows the times for the shell creation and horizon formation; {\bf (b)} in addition illustrates three distinct regimes of radial null geodesics.
(Note that the last (upper right) part of geodesic $III$ is only relevant for nearly-null spacelike geodesics which can bounce off the singularity.
Also, to avoid cluttering the diagram, the starting time $t_i$ and ending time $t_o$ are labeled only for geodesic $I$.)}
\label{collapsePD}
\end{center}
\end{figure}

The Penrose diagram for the collapse is sketched in
\fig{collapsePD}.  There is only one asymptotic
region; below the shell, the spacetime is pure AdS, whereas above,
it is exactly \SAdS.\footnote{The singularity is drawn with a
tilt for more realistic representation of the Penrose diagram in
$d>3$ dimensions (the shape of the singularity depends on $d$ and $r_+$);
however, here the ``Penrose diagram" is to be treated as a sketch for ease of visualisation rather than an exact causal diagram.}
The two important times, which are labeled in \fig{collapsePD} (a),
are the time of the creation of the shell, $t_s$, and the time of
the horizon formation. In global coordinates, the latter occurs in
the bulk at the origin $r=0$ at $t= t_H$; however, this need not
correspond directly to the CFT time, since we're describing an
event at $r=0$ rather than at $r=\infty$.  Instead, the important
CFT time related to the horizon formation is given by $t=t_h$
(labeled on the boundary), which is the time from which a radial
null geodesic would have to start in order to hit the horizon
formation event $(r=0, t=t_H)$.

Since this is a time-dependent geometry, the behavior of a null
geodesic depends on the time $t_i$ of the initial point.
Radial null geodesics therefore separate into three qualitatively distinct classes, indicated in \fig{collapsePD} (b):
\begin{enumerate}
\item[I. ]
For $t_i < t_s - \pi$, the geodesic is given by that of the pure AdS, and $t_o < t_s$.
\item[II. ]
For $ t_s - \pi < t_i < t_h$, the geodesic starts out in AdS, crosses the shell at some point outside the horizon, and continues in \SAdS, reemerging to the boundary at $t_o \in (t_s,\infty)$.
\item[III. ]
For $t_h < t_i < t_s$, the geodesic starts out in AdS but crosses the shell at some point inside the horizon, and therefore hits the singularity (the last part of the geodesic III sketched in \fig{collapsePD} (b) pertains to nearly null spacelike geodesics which can bounce off the singularity).
For $t_i > t_s$, the radial null geodesic starts out in \SAdS\ and therefore crashes into the singularity; whereas for the nearly-null spacelike geodesics, this case is already covered above by reversing the orientation.
\end{enumerate}

We discussed geodesics in AdS and \SAdS\ spacetimes in previous sections; here we match them together to analyse the behaviour in the collapse geometry.  We focus on null geodesics of Type \II, with $t_i \in (t_s - \pi, t_h)$; this will be the interesting regime where we can probe the event horizon formation.
Note that $t_h \in (t_s - \pi, t_s)$, with $t_h$ occurring earlier for larger black hole.
In particular, using ray tracing within  the AdS spacetime
one can show that the horizon formation time is related to the shell creation time $t_s$ and its mass, or equivalently $\rh$, by (see \App{timede} for a derivation)
 \eqn{zeroJth}{
 t_h = t_s - 2 \tan^{-1} r_+ \ .
}

Let us  first consider radial null geodesics. From the Penrose
diagram, it is clear that as $t_i \to t_h$ from below, $t_o \to
\infty$, while for $t_i > t_h$, the null geodesic falls into the
singularity and does not come back to the boundary. Since this will constitute the sharp CFT signature of the horizon formation event in the bulk, let us examine this feature in greater detail. The manner in which $t_o$ diverges as $\delta t \equiv
t_h - t_i \to 0^+$ can be obtained by following Hawking's ray
tracing argument~\cite{Hawking:1974sw}, which gives
 \be \label{delay}
 t_o \approx -{1 \ov \kappa} \log \delta t + {\rm constant}
 \qquad \thus \qquad
 \delta t \sim e^{- \kappa \, t_o}
 \ee
where $\kappa$ is the surface gravity of the black hole.\footnote{ Intuitively Eq.(\ref{delay}) reflects the fact that the redshift experienced by the outgoing null ray increases exponentially with $t_o$ as $e^{\kappa t_o}$ when $t_o \to \infty$.} To see
(\ref{delay}), consider a  point $P_+$ on the horizon just outside
the shell. Let $v^\mu$ to be the vector tangent to the horizon at
$P_+$ and  $u^\mu$ be the null vector directed normally outward from the
horizon normalized so that $u \cdot v =1$.  The vector at $P_+$
which connects the horizon with the null geodesic can thus be
written as $\epsilon \, u^\mu$ for some $\epsilon$. As $t_o \to
\infty$, $\ep$ can be expressed in terms of $t_o$ as
 \be \label{de1}
 \ep \approx A \, e^{-\kappa t_o}, \qquad t_o \to \infty
 \ee
with $A$ some constant independent of $t_o$. Now parallel
transport $u^\mu$ and $v^\mu$ across the shell to the point $P_-$
on the horizon just inside the shell.\footnote{ Note that
$u^\mu$ and $v^\mu$ are continuous when crossing the shell.}  $\ep$ remains
invariant under this transformation. At $P_-$ one can express $\ep$ in terms of $\delta t $
as
 \be \label{de2}
 \ep \approx B \, \delta t
 \ee
with some constant $B$. Comparing (\ref{de1}) and (\ref{de2}), we
obtain the desired relation (\ref{delay}).

Note that the leading behavior (\ie, the logarithmic dependence) in
(\ref{delay}) is rather robust, independent of the collapsing configuration and the type of black hole formed (charged, rotating). In \App{timede} we also give alternative derivation
of (\ref{delay}) for some explicit examples.
While our arguments so far were of a qualitative nature, we can make rather precise predictions for where the singularity in the collapse state correlators should appear.  This is carried out in some detail in \App{timede}, where we calculate $t_o$ in terms of $t_i$ and the parameters describing the state, $t_s$, $\rh$, and $\k$, to be
\eqn{toexpl}{ t_o = t_s + {2 \over r_+ \, \k} \((1+\rh^2) \,
{\pi \over 2}
 - (1+\rh^2) \, \tan^{-1} {r_c \over 1+\rh^2} +
r_+ \, \tanh^{-1} {r_c \over r_+} \)  }
where $r_c$ denotes the crossing radius where the geodesic intersects the shell,
$r_c =  \tan \( {t_s - t_i \over 2} \)$.

To summarize, we find that radial null geodesics are sensitive
probes of horizon formation. The infinite redshift at the horizon
translates into a sharp time scale $t_h$ \zeroJth\ at the
boundary.

Now let us consider non-radial null geodesics. A natural question
is whether there exists an analogous value $t_h (\a)$ for
each\footnote{Since the shell geometry is not static,
$E$ jumps across the shell; in defining $\a$ we use the initial
$E$ \ie, its value in pure AdS.} $\al = E/J$ so that as $t_i \to
t_h (\al)$, the corresponding $t_o \to \infty$. The answer turns
out to be yes for a range of values of $\al$, but for a different
reason from that for the radial geodesics. By the very nature of an
event horizon, we expect that the only null geodesic which truly
samples the event horizon formation is the radial one; any other
null geodesic which approaches the horizon non-radially cannot
escape to the boundary. In \App{vaidyaapp}, we describe how to
compute $t_h (\al)$ for general $\al$. We show that all geodesics
with angular momentum for which $t_o$ diverges take that long to
reach the boundary not because of being trapped near the event
horizon, but rather because of circling around a null circular
orbit at a given finite distance outside the horizon.

\begin{figure}[htbp]
\begin{center}
\includegraphics[width=4in]{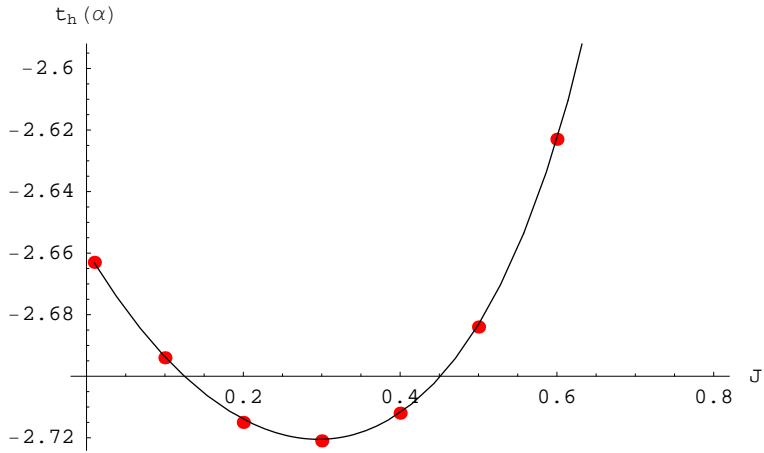}
\caption{Variation of $t_h(\a={1\over J})$  as a function of
angular momentum $J$ in the thin shell spacetime \metricgen\
(solid line). The intersection of the curve with vertical axis
gives $t_h$, while the minimum gives $t_c$. Red dots are data
extracted from \fig{vadtito}, which were obtained by numerical
integration of the geodesics in the Vaidya spacetime (\ref{vaidyamet}), (\ref{vaidyf}) with $v_s =0.001$.
  }
\label{vadtj}
\end{center}
\end{figure}

 It is easy to see that
$t_h(\a)$ depends non-trivially on $\a$; the presence of angular
momentum causes the geodesic to sample a different region of the
spacetime geometry. In the limit $\a \to 1$ (maximal angular
momentum) we expect that the geodesic stays arbitrarily close to
the boundary, so that it doesn't sample the black hole geometry,
and in particular there can be no divergence in $t_{o}$ for any
finite $t_i$. In the opposite limit of zero angular momentum, we
have already seen that $t_h(\infty)$ is given by the expression
for $t_h$ for the radial geodesic \zeroJth. The result of
\App{vaidyaapp} is summarized in Fig. \ref{vadtj}, which plots
$t_h (\a)$ as a function of $J \sim 1/\a $. In particular, we note that:
 \eqn{talphapred}{\eqalign{
t_h(\a) &\; \to \; t_h = t_{s} -2 \, \tan^{-1}(\rh) \ , \qquad  \a
\to \infty \cr
 t_h(\a) & \; \to \; t_s \ , \qquad \a \to \a_0
  }}
while for $1< \al < \al_0$, $t_o$ never diverges, where $\al_0$ was
introduced in~(\ref{deal}). Recall that the \SAdS\ geometry
contains a null circular orbit with $\al=\al_0$ at $r = \sqrt{2\, 
\mu}$.
 As Fig. \ref{vadtj} demonstrates, $t_h(\alpha)$
 is not a monotonic function of $1/\a$. In particular, the minimal
 value of $t_h (\al)$, which we will denote as $t_c$, is smaller
 than $t_h$, the horizon formation time. $t_c$ is the boundary time scale
 at which the null circular orbit of the newly formed black hole geometry
 is first probed.

\begin{figure}[htbp]
\begin{center}
\includegraphics[width=6.5in]{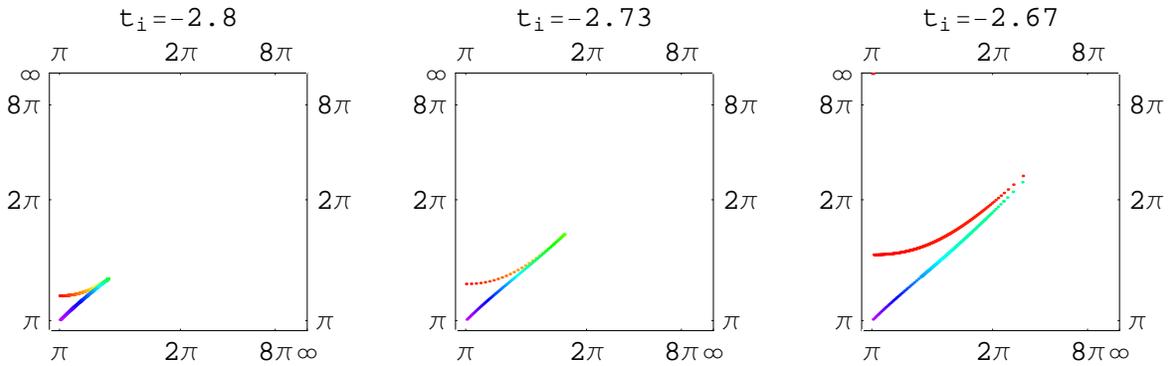}
\caption{Plot of  $(\D \ph,\D t) $ for 3 values of $t_i$ in the
thin shell geometry (same configuration as in \fig{vadtj} and
\fig{vadtito}(a)), for various values of $\a \in  (1,\infty)$. The
horizontal and vertical axes show $\D \ph$ and $\D t$
respectively, rescaled by $\tan^{-1}$ to fit the full plane. The
left endpoint of the upper branch corresponds to $\al=\infty$
(radial geodesic), while the left endpoint of the lower branch
corresponds to $\al =1$. (a) $t_i = -2.8 < t_c \approx -2.72$; (b) $t_i = -2.73$
only slightly smaller than $t_c$; (c) $t_i = -2.67$  slightly smaller than
$t_h \approx -2.66$.   Here $t_c$ is the minimal value of
the curve in \fig{vadtj} and $t_h$ is intersection of the curve in \fig{vadtj}
with the vertical axis. As $t_i \to t_c$ the cusp should reach
$(\infty, \infty)$ (upper right corner) and as $t_i \to t_h$, the
left endpoint of the upper branch should reach $(\pi, \infty)$
(the upper left corner). Note that as apparent from \fig{vadtito},
the endpoints vary rather sharply with $t_i$ for $t_i \approx
t_h(\a)$, so it is difficult numerically to sample the large
values of $\D \ph$, $\D t$.} \label{colltph}
\end{center}
\end{figure}

In \fig{colltph}, we plot the endpoints of various geodesics on
the $\D t - \D \ph$ plane\footnote{Since $\a$ is the only
continuous free parameter describing the null geodesics, this will
produce a curve of endpoints on the $\D t - \D \ph$ plane.} for a
fixed initial time $t_i \in (t_s-\pi, t_s)$. Note that in contrast
to the analogous plots in earlier sections for other geometries,
the background here does not have a time translational symmetry.
For ease of visualization we unwrap the $\ph$ direction and at the
same time compactify both the $\D t$ and $\D \ph$ so that we can
examine the full $\D t - \D \ph$ plane.  For each fixed $t_i$, we
plot the end-point curve, color-coded by $\a$. For $t_i < t_s -
\pi$, the endpoints would all clump into the single point $(\D
\ph,\D t) = (\pi,\pi)$.  As we increase $t_i$ so that the
geodesics start sampling the shell, the endpoints begin to spread
in the manner shown on the left in \fig{colltph}, into a cusp
similar to the star geometry (\cf\ \fig{startphrho}).  As $t_i$
reaches the minimum of the curve $t_h(\a)$ shown in \fig{vadtj}
(\ie, $t \to t_c$), the cusp extents to $(\D \ph,\D t) =
(\infty,\infty)$ (upper right corner of the plots) with
 \be \label{rrr}
  \D t (\al) \approx {1 \ov \am} \, \D \ph (\al) \to \infty \ .
 \ee
Increasing $t_i$ further, the cusp still reaches $(\D \ph,\D t) =
(\infty,\infty)$ with (\ref{rrr}), but now for two different
values of $\a$, given by solution to $t_h(\a) = t_i$. At the same
time, $\D t$ for the radial geodesic increases, diverging as $t_i
\to t_h$. This will appear in \fig{colltph} as the left end point
of the upper branch moving to $(\pi, \infty)$. The right plot in
\fig{colltph} is taken for $t_i$ slightly smaller than $t_h$. Note
that equation (\ref{rrr}) follows from the fact that as $t_i \to t_h
(\al)$, the geodesic goes around the circular null orbit with a
period ${2 \pi \ov \al_0}$ infinitely many times.

One can generalize the sharp shell to a smeared-out version which
is physically more realistic. To that end, we consider a Vaidya
spacetime of the form 
\eqn{vaidyamet}{ ds^2 = - f(r,v) \,  dv^2 +
2 \, dv \, dr + r^2 \, d \Om^2_3 } 
with $f(r,v)$ smoothly
transitioning between (\ref{fin}) at $v \to - \infty$ and
(\ref{fout}) as $v \to + \infty$; a convenient form to use is
\eqn{vaidyf}{ f(r,v) = r^2 + 1 - {\mu \over r^2} \( {1+ \tanh{v
\over v_s} \over 2} \) }
 In particular, the shell is inserted at $v=0=t_s + \pi/2$, and has `thickness' $v_s$.
 As $v_s \to 0$, we recover the collapse spacetime (\ref{metricgen})
 written in ingoing coordinates.

\begin{figure}[htbp]
\begin{center}
\includegraphics[width=7in]{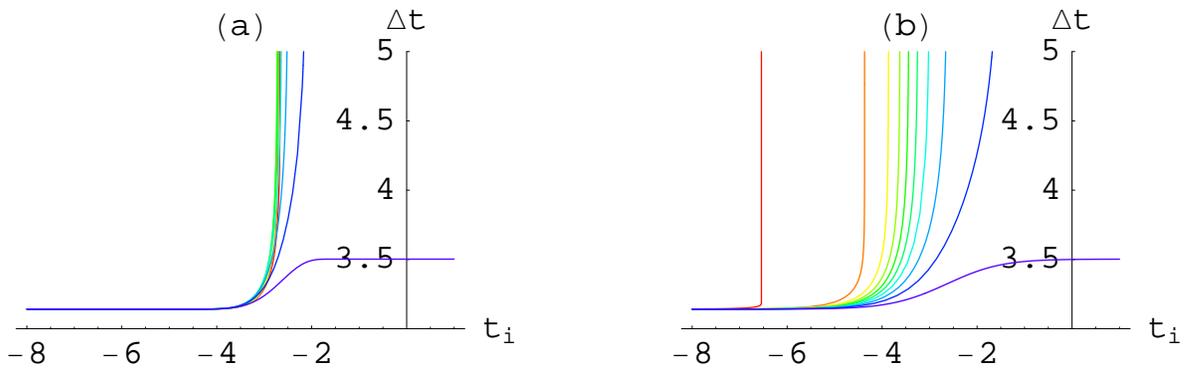}
\caption{Plot of $\Delta t$ as a function of $t_i$ for non-radial
null geodesics in the Vaidya metric with the effective
``thickness" of the shell given by {\bf (a)} $v_s = 0.001$ and
{\bf (b)} $v_s = 1$.  The various curves differenct angular
momenta, $J= 0.01, 0.1, 0.2, 0.3, \ldots$. }
\label{vadtito}
\end{center}
\end{figure}
\fig{vadtito} shows the value of $\Delta t=t_o-t_i$ for a set of
non-radial geodesics in the spacetime (\ref{vaidyamet}) with
various values of $\alpha$, all starting at the initial time
$t_i$; in (a) the shell is very thin whereas in (b) the shell is
considerably smeared out.   The value of $t_h(\alpha)$ can be read
off from the plot; it corresponds to the value of $t_i$ at which
$\D t$ diverges for a geodesic with that $\a$.  For the thick
shell in \fig{vadtito} (b), we see that these times are
well-separated, and increase monotonically with increasing $J/E$
(decreasing $\a$).

This illustrates that the sharp signature of the horizon formation event is not 
limited to the thin-shell collapse (bulk spacetimes with a null
shell sharply separating pure AdS from \SAdS):  For any spherically
symmetric spacetime in which an event horizon forms at a time
$t_H$ at the origin, reachable by an ingoing radial null geodesic
starting at the boundary at $t_h$, we can find $t_h$ by exactly
the same method. In particular, our method constitutes following
the red (leftmost) curve in \fig{vadtito} (b).

To summarize,  whereas only the radial null geodesic starting at
$t_h$ samples the horizon formation event, the non-radial null
geodesics provide further details about the geometry, in
particular the null circular orbit of the newly formed black hole
geometry. Furthermore the  variation of the curves $\D t (\D \ph)$ with respect to $t_i$ provides spatio-temporal information both dynamically around the horizon formation time and spatially in the vicinity of the horizon.

\subsection{Gauge theory signatures}

We can now translate the behaviour of null geodesics in the bulk
collapse spacetime to predictions regarding boundary correlation
functions. Our discussion below should apply to generic highly
excited states in the boundary theory\footnote{By highly excited
states we mean states of energies of order $cN^2$, with
$N$-independent constant $c$ sufficiently big. By generic states
we mean states which are generic superpositions of energy
eigenstates. Note that energy eigenstates do not give rise to the
desired time dependence.}.

Consider the CFT correlators of the form $\expval{\CO(t_i,\Om) \,
\CO(t_o,-\Om)}_{\rm shell}$ evaluated explicitly in the state of
the shell. Modelling the shell as being created by an operator
$\CS$ inserted in the boundary at $t =t_s$,  we have
\eqn{shellprobe}{\eqalign{ \expval{\CO(t_o,\Om') \,
\CO(t_i,\Om)}_{\rm shell} & = \bra{0} \, \CO(t_o,\Om') \,
\CO(t_i,\Om) \, \ket{0}  \ , \qquad  t_i < t_o < t_s \cr & =
\bra{0} \, \CO(t_o,\Om') \, \CO(t_i,\Om) \, \CS (t_s)\, \ket{0} \
, \qquad t_i  < t_s < t_o \cr & = \bra{0} \, \CS^{\dagger} (t_s)
\CO(t_o,\Om') \, \CO(t_i,\Om) \, \CS (t_s) \, \ket{0}  \ , \qquad
t_s<t_i < t_o }}
By our previous arguments, we would expect this correlator to
become singular when the insertion points are connected by a null
geodesic. More explicitly,
\begin{enumerate}
\item[I. ]
  When  $t_o < t_s$, the correlator is simply the standard vacuum two point
  function. The singularities are given by the usual light cone
  singularities of the boundary theory. As is clear from geodesic I in
  \fig{collapsePD}(b)), all null geodesics lie entirely in the AdS
  region.
\item[II.  ]
  When $t_i < t_s < t_o$, the correlator is given by the second
  line of \shellprobe. From our discussion of null geodesics in last subsection,
 the pattern of bulk-cone singularities for this correlator has a rich
 structure, which reflects the formation of the event horizon and the null circular orbit
 in the bulk.

 Suppose experimentalists in the boundary theory are
 able to measure the observable $\expval{\CO(t_i,\Om) \, \CO(t_o,-\Om)}_{\rm
 shell}$ for all $t_i$ and $t_o$ satisfying $t_i < t_s < t_o$.
 Then by carefully plotting the locations of singularities of the correlator in
 the $\D t - \D \varphi$ plane for a given $t_i$, they should
 recover various plots of \fig{colltph}. In particular two distinct time scales
 emerge by comparing the pattern of singularities for different
 $t_i$. The first is the time $t_c$ when the cusp in \fig{colltph}
 reaches $(\infty, \infty)$. This is the time scale that non-radial
 null geodesics originating from the pure AdS geometry start being
 trapped by the circular orbit of the newly formed black hole. The
 slope of the line reaching the infinity gives the period of the
 orbit. The second is the time $t_h$ that the left end of the
 upper branch (which corresponds to radial geodesics)
 reaches $(\pi,  \infty)$. This is the time of horizon
 formation as probed by a radial null geodesic. The bulk
 time of horizon formation in global AdS coordinates is given by
 $t_H = t_h + {\pi \ov 2}$.

 \item[III. ] When $t_s < t_i< t_o$, the correlator is given by the
 third line of \shellprobe. In this case, all singularities of the
 correlator are the same\footnote{The correlator itself does not necessarily coincide
 with a thermal correlation function. Only the singularity structure does.}
 as those of a finite temperature
 correlation function with the temperature $T$ given by
 \be
 {1 \ov T} = {\p \log \Om (E) \ov \p E}
 \ee
where $E$ is the energy of the shell state and $\Om (E)$ is the
density of states of the CFT. This follows from that all type \II\
geodesics with $t_s < t_i< t_o$ are identical to those of an AdS
black hole.

\end{enumerate}

In the above we discussed the pattern of bulk-cone singularities
of \shellprobe. There might also be more subtle signals, on a
secondary sheet (so we would find the singularity in the
correlator only after a suitable analytic continuation) when the
insertions are separated by a spacelike geodesic which is
arbitrarily close to being null as illustrated for the eternal
black hole singularity in \cite{Fidkowski:2003nf}. In other words,
the corresponding geodesic passes through the black hole and
bounces off the singularity\footnote{However, note that there is
no longer a single radial null
  geodesic connecting the insertion points.}, see e.g. geodesic III
of~\fig{collapsePD}(b). Such a situation could arise when $t_i >
t_h$ (which, if we define $t_i < t_o$, also implies that  $t_i <
t_s < t_o$). We will discuss it in more detail in 
\sec{bhsing}.

\subsection{Signature of the black hole singularity?}
\label{bhsing}

\begin{figure}[htbp]
\begin{center}
\includegraphics[width=6in]{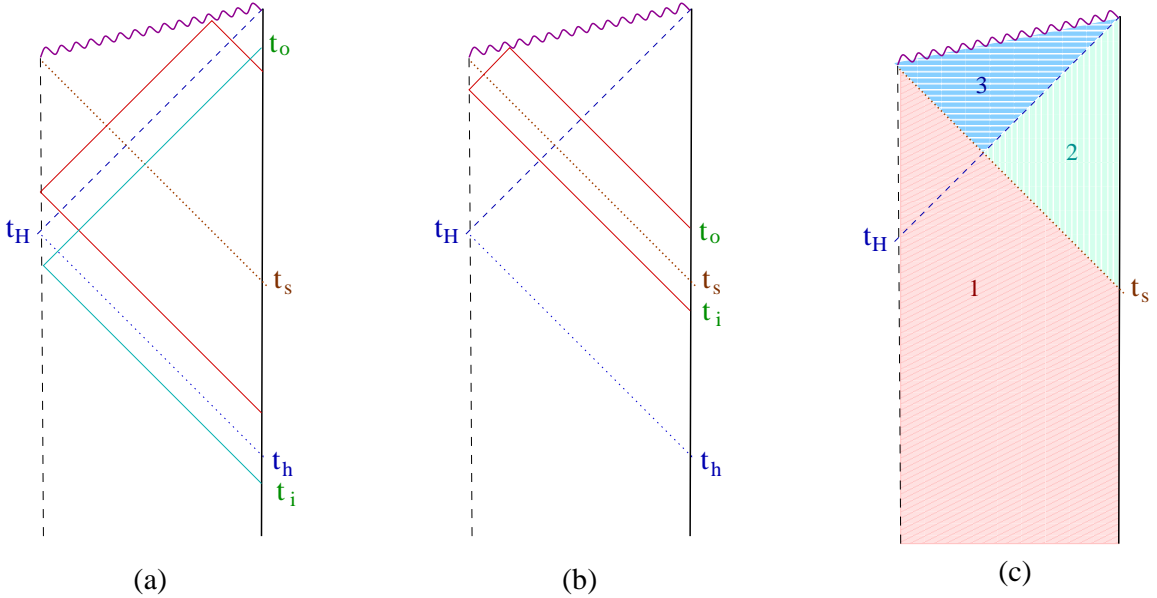}
\caption{Penrose diagram for a collapsed black hole}
\label{collapsePDs}
\end{center}
\end{figure}

In addition to the null geodesics considered in
\fig{collapsePD}, it is also of interest to consider some
spacelike geodesics. In particular, spacelike geodesics have the
important feature that they `bounce off' the black hole
singularity~\cite{Fidkowski:2003nf}, and thus could provide
valuable information about the black hole singularity. There are
two other types spacelike geodesics in the collapse background
which are of particular interest (see \fig{collapsePDs}):
\begin{enumerate}
\item $ t_i \to t_h$ from above and $t_o \to \infty$. As can be
seen from Fig.\ref{collapsePDs}a, the geodesic passes inside the
horizon and bounces off the singularity.  We will calculate below
the dependence of $t_o$ on $\delta t = t_h - t_i$.
\item As $t_i \to t_s$,  $t_o \to t_s.$ The corresponding geodesic is
sketched in Fig.\ref{collapsePDs}b. One expects that $t_o - t_s
\sim (t_s - t_i)^{\gamma}$, where $\gamma$ should depend on
how strongly geodesics are repelled from the singularity.
\end{enumerate}
Note that spacelike geodesics are parameterized by $E$ and $J$, but here we set $J=0$ to focus on radial geodesics, and moreover we take the infinite $E$ limit, so that
their trajectories converge to the radial null geodesics, except for the
important point that they `bounce off' the black hole singularity.
As before, we can find geodesics in the full collapsing geometry by patching the geodesic segments in the respective regions; here in addition to AdS (region 1 in \fig{collapsePDs}c) and the \SAdS\ region outside the horizon (region 2 in \fig{collapsePDs}c), we now have to consider \SAdS\ inside the horizon as well (region 3 in \fig{collapsePDs}c).  This is presented in detail in \App{timede}; here we only quote the main results.

\begin{enumerate}
\item As $ t_i \to t_h^+$, $t_o \to \infty$ in the same manner as for $ t_i \to t_h^-$ (\cf\ Eq.(\ref{delay})), namely:
\eqn{delayin}{
-\delta t = t_i - t_h \sim e^{-\kappa \, t_o}}
\item Let $t_s -t_i \equiv \eps$. Then expanding $t_o - t_s$ to third order in
$\eps$, we find that
\eqn{}{t_o - t_s \sim \eps^3 \ . }
with a coefficient which depends on the parameters of the final black hole -- see (\ref{gammascaling}).
\end{enumerate}

As argued around (\ref{delay}), it is clear that the behaviour \delayin\ is robust and is expected to hold in any spacetime dimension. The behaviour of the geodesic which bouces off the singularity in  \fig{collapsePDs}b, and in particular the exponent $\gamma$, depends crucially on the dimension; $\gamma= d-2$ in AdS$_{d}$. Intuitively this is due to the fact that the curvature of the singularity increases with spacetime dimension.

\section{Discussion}
\label{discuss}

We have argued that the AdS/CFT correspondence implies the
existence of ``bulk-cone singularities'' in boundary theory
correlation functions, which lie inside the light-cone. The pattern
of singularities can in turn be exploited to extract information
about the bulk geometry.
 Our argument revolved around the essential
point that Green's functions for quantum fields in curved space
are singular when the arguments are null separated. While the
geodesic approximation was necessary to derive the conclusion, we
should point out that one expects the result to be true as long as
it makes sense to talk about the spacetime geometry. In other words, we
expect validity in $\a'$ expansion, but not necessarily in 
$g_s$ expansion (where it is plausible to have a breakdown of the
geometric picture).

We have demonstrated the pattern of singularities in the field
theory correlators for several distinct scenarios:  in radiation star, eternal \SAdS\ black hole, and the geometry of a collapsing shell.  We have seen that small deformations of bulk geometries
can in principle produce significant differences in the structure of singularities which are discernible on the boundary. In this
context it is also satisfying that the vacuum state of the CFT, corresponding to
pure AdS spacetime, has a distinguished role, as evidenced by the
theorem of \cite{Gao:2000ga}. Further, we are guaranteed that
bulk-cone singularities for correlators evaluated in 
excited states occur at later times compared to the vacuum correlators, as necessitated by causality.

The novel application of our proposal was to the case of dynamical
spacetimes. We concentrated on the simple scenario of a null shell
collapse in AdS/CFT. One of the fascinating aspects of the singularity
structure of the correlations is that they encode the spacetime
event corresponding to event horizon formation in the bulk
unambiguously. The characteristic signature we discussed is quite
robust and can be easily distinguished from other singularities in
the correlation functions. Further, the detailed pattern of
singularities carries information about the geometry of the black hole
that is formed. By using the non-local nature of the correlation
functions we are able to extract information about the details of the
bulk spacetime, while, in contrast to the recent discussions of
probing the black singularity in AdS/CFT
\cite{Fidkowski:2003nf,Festuccia:2005pi}, we are not reliant on
analytic properties of the CFT correlators.

It would be very desirable to test our predictions of 
the existence of the bulk-cone singularities directly within the
boundary theory. Note that the classical limit corresponds to the
large $N$ and large 't Hooft coupling limits of the boundary
theory. It would be interesting to understand whether the
appearance of bulk-cone singularities for correlation
functions in excited states is an artifact of the large $N$ and/or 
large 't Hooft coupling limits, or  a generic feature for field
theories on compact spaces.\footnote{Note that bulk-cone singularities 
do not arise for non-compact boundary as shown in \App{apppoincare} for states preserving the boundary isometries.} Further, we expect that the semiclassical picture that we used in the bulk to be valid only in the large $N$ limit -- at any finite $N$ we will have contributions from other semi-classical geometries that are sub-dominant saddle points of the quantum gravity path integral. In these circumstances it is not clear whether there is a precise meaning to the bulk light-cone. Nevertheless, we should note that the bulk-cone singularities discussed here are eminently trackable in the field theory even beyond the planar limit of large $N$. Understanding the role of these in some geometric terms is an interesting challenge. 

An intriguing avenue to explore is to extend these considerations
to ``microstate geometries'' proposed for black hole spacetimes. In
situations where we have  an explicit map between the spacetime
geometries and the field theory states such as  two charge D1-D5
system \cite{Mathur:2005zp,Mathur:2005ai}, the 1/2 BPS sector of
$\CN =4$ SYM \cite{Lin:2004nb}, one can ask whether the pattern of
singularities differs significantly between different states.  The
general arguments regarding the correlation functions of typical
states that comprise the black hole entropy in
\cite{Balasubramanian:2005kk,Balasubramanian:2005mg,Balasubramanian:2005qu,Balasubramanian:2006jt}
would seem to suggest that the structure of the singularities is
not capable of discerning the fine distinctions between the
geometries. This is however counter to the gravitational
intuition: naively one expects geometries that differ from each
other to have differing behaviour of Green's functions, especially
with regard to the location of the singularities. A potential
resolution of this discrepancy is that the geometric picture is
incorrect for the ``microstate geometries'' as one has to consider
a suitable wavefunction over the quantum moduli space of
solutions. This is an interesting question that deserves to be
explored further.

Another interesting generalization of our set-up would be to model the
AdS/CFT analog of Choptuik scaling \cite{Choptuik:1992jv}. Given
that we have a specific prediction for the field theory signature
of black hole event horizon formation, one can ask whether the
critical behaviour observed during the collapse has a field
theoretic image and map out the details of the correspondence in
this case.

\section*{Acknowledgements}
We would like to thank Allan Adams, Dan Freedman, Gary Horowitz,
Bob Jaffe, Matt Kleban Albion Lawrence,  Don Marolf, Simon Ross, Steve Shenker, Amit Sever for discussions. We would also like to thank KITP, Santa Barbara
for kind hospitality during the ``Scanning New Horizons: GR beyond
4 dimensions" program where this project was initiated. VH and MR
would also like to thank the Aspen Centre for Physics for
hospitality during the concluding stages of this project. HL is
supported in part by the A.~P.~Sloan Foundation, the U.S.
Department of Energy (DOE) OJI program and by the DOE under
cooperative research agreement \#DF-FC02-94ER40818.

\pagebreak
\startappendix
\Appendix{Pure AdS}
\label{appAdS}

For ease of comparison of our results on the star and other
geometries in what follows, we first summarise the basic facts
regarding geodesics in AdS, concentrating on null and spacelike
geodesics.

\subsection{Geodesics in AdS: summary}

Consider $d+1$-dimensional AdS in global coordinates.  The metric
can be written as
\eqn{adsmet}{\eqalign{ ds^2 &= -f(r) \, dt^2 + {dr^2 \over f(r)} + r^2 \,
d\Om_{d-1}^2 \cr
f(r) &= 1+r^2  \ . }}
The geodesic equations are as in \geodbhs, with $f(r)$ being
that given in \adsmet.

As we are primarily interested in the endpoints of null or spacelike geodesics (ultimately with diverging regularised proper length).  To that end, the useful quantities are the temporal and angular separation of the geodesic endpoints, which we denote by $\D t$ and $\D \ph$, respectively.  Let us first focus on spacelike geodesics, since as we will see, these limit to null geodesics.  We can integrate the geodesic equations to obtain:
\eqn{Dtspace}{ \Delta t = {\pi \over 2} +  \sin^{-1} \left( { E^2 -J^2 -1 \over e^{-L} } \right) }
\eqn{Dphspace}{  \Delta \ph = {\pi \over 2} +  \sin^{-1} \left( { E^2 -J^2 +1 \over e^{-L} } \right) }
\eqn{Lenspace}{  e^{-L} = \sqrt{ (E^2-J^2)^2 + 2 (E^2 + J^2) +1} }
where $L$ is the geodesic length after regularising away the divergent piece $\log R_C^2$ ($R_C$ is the radial cut-off).
Finally, another useful expression is the minimum $r$ value reached by the geodesic $r_{min}$
\eqn{}{  2 \, r_{min}^2 = - (E^2 -J^2 +1) + e^{-L}}

From the expression for $e^{-L}$ it is clear that we have a large negative length when
\begin{enumerate}
\item  $E \rightarrow \infty$  with  $J$ fixed  $\thuss L \sim - \log E^2$
\item $J \rightarrow \infty$ with $E$ fixed $\thuss  L \sim -\log J^2$
\item  $J \sim E \rightarrow \infty$  $\thuss  L \sim -\log E$
\end{enumerate}

Now consider two boundary points $A$ and $B$ are separated by boundary coordinate distance $\D t = \pi - \epsilon_t$ and $\D \ph = \pi -\epsilon_{\ph}$. We can easily see that for any spacelike separation $0 < \epsilon_{t,\ph} < \pi$, there is a unique geodesic joining $A$ and $B$; specifically we can calculate $E$ and $J$ for the geodesic given $\epsilon_{t,\ph}$ :
\eqn{}{  E = {\sin \epsilon_t \over \cos \epsilon_{\ph} - \cos \epsilon_t} }
\eqn{}{  J = {\sin \epsilon_{\ph} \over \cos \epsilon_{\ph} - \cos \epsilon_t} }
and its length is
\eqn{}{  L = - \log 2 + \log ( \cos \epsilon_{\ph} - \cos \epsilon_t) }
%
\subsection{Spacelike \vs\ null geodesics}

Null geodesics in AdS only connect points with $\D t = \D \ph =
\pi$, irrespective of the parameter $E/J$.  While this is a well
known fact for radial null geodesics (for which $J=0$), it is true
that even geodesics carrying angular momentum emerge at the
anti-podal point on the sphere in exactly AdS time. As mentioned
in \sec{nullgsing} these null geodesics endpoints being coincident
with the endpoints of purely boundary null geodesics, we do not
see extra singularities in the correlation functions.

A naive puzzle about the null limit arises when we consider the
difference between the angular separation of the endpoints $\D
\ph$ for spacelike and null geodesics.  Spacelike geodesics in AdS
allow any $0 \le \D \ph \le \pi$; indeed for $E=0$ (constant $t$
slice) the geodesics plotted on the Poincare disk are circular
arcs with diameter related to the angular momentum. On the other
hand, null geodesics always reemmerge at the anti-podal point $\D
\ph = \pi$.  How do the geodesics change their behaviour as we
take $E \to \infty$ at fixed $J$?
\begin{figure}[htbp]
\begin{center}
\includegraphics[width=7.5in]{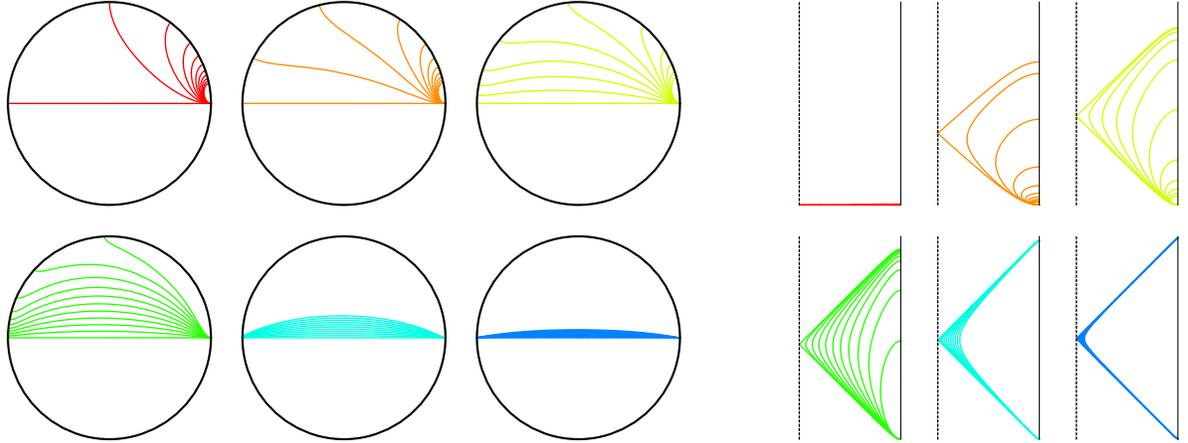}
\caption{Spacelike geodesics in AdS, projected to a constant $t$
slice (left) and the $t-r$ plane (right),
for six different values of energy ($E = 0,2,5,10,30,80$).
For each $E$, geodesics with different values of angular momentum
($J= 0,1,2,\ldots,10$) are plotted.  The bold circles represent
the boundary at $\tan r = {\pi \over 2}$. The dashed vertical lines
correspond to the origin $r=0$ and the bold vertical lines to the
boundary at $\tan r = {\pi \over 2}$.  The range of $t$ plotted is
$(0,\pi)$.} \label{AdSgeodsphrt}
\end{center}
\end{figure}
\fig{AdSgeodsphrt} illustrates this
limit.  For any fixed finite $J$, as $E \to \infty$, we see that
the geodesic indeed converges to the radial null geodesic and $\D
\ph \to \pi$.

Nevertheless, the full set of singularities of the boundary
correlation function \adscorr\ is larger than those captured by
the points with $\D t = \D \ph = \pi$. It is interesting to ask
whether there are any bulk geodesics (with divergent regularized proper length)
 that connect points with $\D t = \D \ph \neq \pi$. From \Lenspace, one  concludes that the regularised proper length along a spacelike geodesic diverges not only when $E \to \infty$, but also when $J \to \infty$.  Does this mean that this is an equally good null limit, and that therefore $\D \ph = \pi$?  The answer is no: as we can see from \Dphspace , as  $J \to \infty$ at fixed $E$, $\D \ph \to 0$.  The geodesic has vanishing proper length because it is so short rather than because its tangent vector is null.  On the boundary, this corresponds to the usual divergence of correlator of operators inserted at the same point.

However, it turns out that there nevertheless are nontrivial almost-null spacelike
geodesics with $\D t = \D \ph \neq \pi$. Consider spacelike geodesics
in AdS parameterized by $E$ and $J$ with the following constraint:
\eqn{}{J^2 = E^2 - \sig \, E}
for some real number $\sig$. Upon taking the limit $E \to \infty$,
we obtain
\eqn{Dtlim}{ \Delta t \to {\pi \over 2} +  \sin^{-1} \( {\sig \over
\sqrt{\sig^2 + 4} } \) }
\eqn{Dphlim}{  \Delta \ph \to {\pi \over 2} +  \sin^{-1} \( {\sig
\over \sqrt{\sig^2 + 4} } \) }
Clearly, these are equal, and take the full range $0 <\D t = \D
\ph < \pi $ for $-\infty < \sig < \infty$. Further, we can check
that this family of geodesics indeed converge onto the boundary
null geodesics.
\begin{figure}[htbp]
\begin{center}
\includegraphics[width=5in]{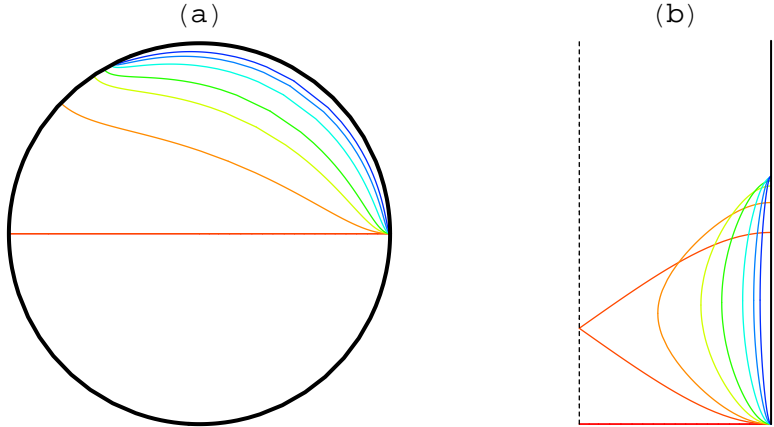}
\caption{Spacelike geodesics in AdS, projected to {\bf (a)} a
constant $t$ slice and {\bf (b)}  the $t-r$ plane, for increasing
values of energy and $J^2 = E^2 - \sig \, E$ with $\sig = 1$ (in the
infinite $E$ limit this would yield $\D t = \D \ph \approx {\pi
\over 3} $).  In  {\bf (a)} the boundary is indicated as the bold
circle. In {\bf (b)} the boundary at $\tan \, r = {\pi \over 2}$
is in bold  and the dashed vertical line is the origin. }
\label{Adsgeodsalim}
\end{center}
\end{figure}
\fig{Adsgeodsalim} demonstrates these features.  We see
that as $E$ increases, the geodesics accumulate at a finite value
of $\D t$ and $\D \ph$, in this case $\sig$ being 1, ${\pi \over 2} +
\sin^{-1} \( {1 \over \sqrt{5} } \) \approx {\pi \over 3} $.

To summarise this discussion, in pure AdS$_{d+1}$ spacetime, null geodesics connect points on the boundary which are anti-podally located on the $\S^{d-1}$ at a time separation $\Delta t = \pi$. Spacelike bulk geodesics on the other hand connect points with $\Delta t = \Delta \ph \neq \pi$ as illustrated by \Dtlim\ and \Dphlim, respectively
(or more generally, \Dtspace\ and \Dphspace).
 
 \newpage
\Appendix{Radiation star in AdS}
\label{appstar}

In this Appendix, we construct the geometry corresponding to a gas of radiation in AdS and study its properties.  As sketched out in \sec{statgeom}, constructing the solution entails specifying the form of the stress-energy-momentum tensor and solving the Einstein's equations.\footnote{
Analogous calculations in 4 dimensions were discussed in \cite{Page:1985em}, with similar qualitative results as we obtain below.}
After filling in the gaps in the presentation of \sec{statgeom}, we extend our study of the star geometry by analysing spacelike geodesics, and then discuss the observable differences between the star geometry and that of a black hole.

\subsection{Construction of the star-AdS spacetime}

As discussed in \sec{statgeom}, one can construct a simple model for a static, spherically symmetric, asymptotically AdS ``star" geometry by solving Einstein's equations with a negative cosmological constant and matter given by a perfect fluid stress tensor corresponding to radiation. The metric coefficients are determined by solving the Einstein's equation
\eqn{Eeq}{
G_{ab} + \Lambda \, g_{ab} = 8 \pi  G_{5} \, T_{ab}
}
where for convenience we will set $8 \pi  G_{5}  \equiv 1$ and $\Lambda = -6$ to set
AdS radius to unity.

The symmetries constrain the metric to take the form \starmet. One can then infer the equations \hstar\ and \mstar\ from the $tt$ component of the Einstein tensor. The $rr$ component of the Einstein tensor yields
\eqn{fratiostar}{
{f' \over f} = { 2 \over r } \, \( {
 {r^2 \over R^2} + {m(r) \over r^2} +{1 \over 12} \, \rho(r) \, r^2 \over
 {r^2 \over R^2} + 1 - { m(r) \over r^2} } \)
}
which can be easily integrated to get $f(r)$ in terms of $m(r)$ and $\rho(r)$.
Finally, using the angular part of Einstein's equation, or equivalently the stress tensor conservation, we can write (for general equation of state)
\eqn{Pstar}{
{d P \over dr} = - (\rho + P)  \, {1 \over 2} \, {f' \over f}
}
This can then be used to derive the system of coupled first order ODEs for $m(r)$ and $\rho(r)$ given by (\ref{m1}) and (\ref{p1}), respectively. The equation for $f(r)$, \fratiostar, can be simplified to \afstar\ using the equations for $m(r)$ and $\rho(r)$.

\subsection{Geodesics in the star geometry}

In \sec{statgeom} we examined some aspects of the star geometry by focusing on null geodesics; here we extend this analysis to spacelike geodesics as well.  While these do not lead to singularities of the boundary correlators, they nevertheless reveal interesting points.
\begin{figure}[htbp]
\begin{center}
\includegraphics[width= 7.5in]{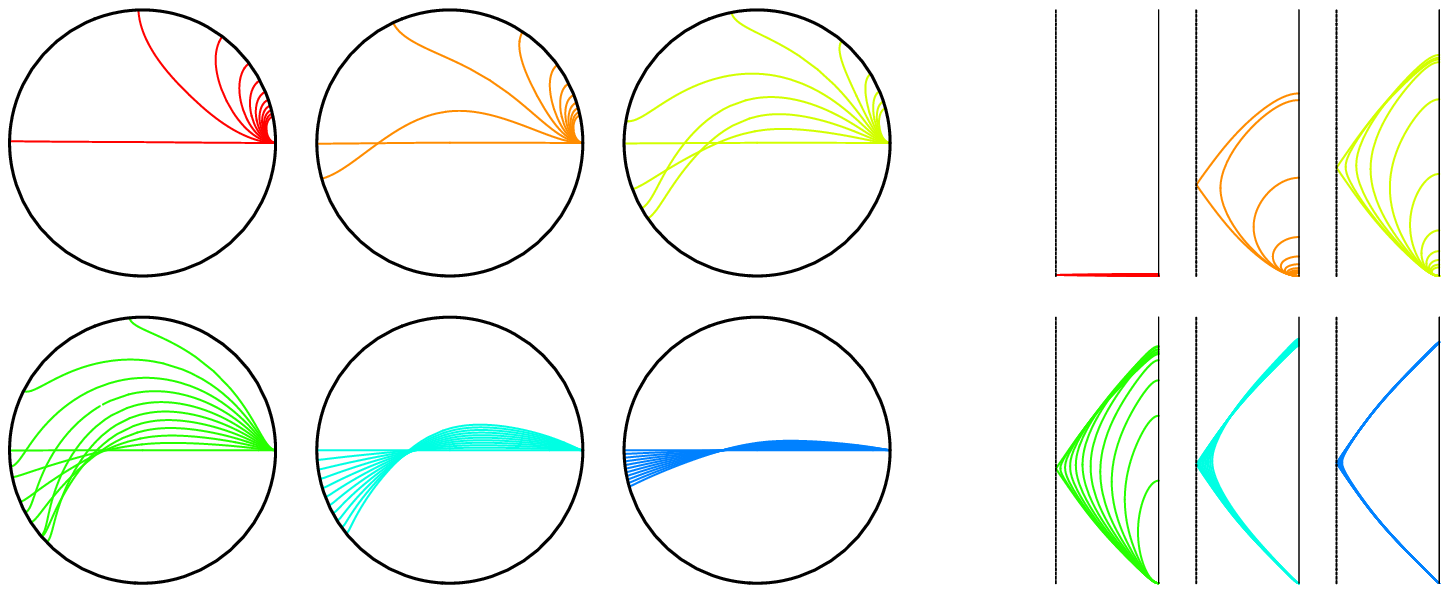}
\caption{Spacelike geodesics in star with $\ro=10$ in AdS,
projected to a constant $t$
slice (left) and the $t-r$ plane (right),
for six different values of energy ($E = 0,2,5,10,30,80$).
For each $E$, geodesics with different values of angular momentum
($J= 0,1,2,\ldots,10$) are plotted.  The bold circles represent
the boundary at $\tan r = {\pi \over 2}$. The dashed vertical lines
correspond to the origin $r=0$ and the bold vertical lines to the
boundary at $\tan r = {\pi \over 2}$.  The range of $t$ plotted is
$(0,1.1 \, \Delta t_0)$. }
\label{stargeodsphrt}
\end{center}
\end{figure}
Since the star in AdS can be vied as a deformation of the pure AdS geometry, it is particularly interesting to contrast the behaviour of spacelike
geodesics in the star geometry with that in AdS. In
\fig{stargeodsphrt}  we show spacelike
geodesics projected onto the constant $t$ and $t - r$ plane
 (these plots are to be compared with \fig{AdSgeodsphrt}, where same conventions
apply). As for null geodesics, we see a focusing  effect, created
by the star in the center of AdS.

\begin{figure}[htbp]
\begin{center}
\includegraphics[width=5in]{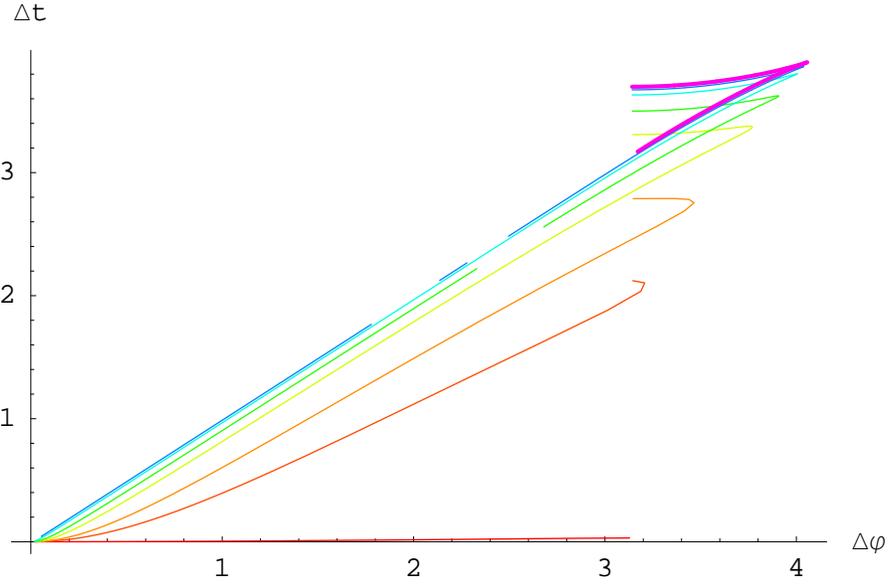}
\caption{Endpoints of spacelike and null geodesics in the geometry
of star with $\ro=10$ in AdS.  Spacelike geodesics for six
different values of energy ($E = 0,2,5,10,30,80$) and a null
geodesic (thick pink curve) is plotted for varying values of $J$
-- varying $J$ traces out  curves in the $\Delta t  -
\Delta \ph$ plane; the top of the curves corresponds to small
values of $J$, while in the large $J$ limit, $\Delta t$ and
$\Delta \ph$ both approach their starting value, namely 0.}
\label{startphinf}
\end{center}
\end{figure}

To see how the endpoints of these various geodesics compare, let
us consider the values $\Delta t$ and $\Delta \ph$ for the
various geodesics. \fig{startphinf} summarizes this. One interesting feature
to note is that since the curves for various energies intersect
each other, it is no longer true that for {\it any} two
spacelike-separated points, there is a unique spacelike geodesic
connecting them.  In particular, unlike the pure AdS case, there
is an open set of endpoints which can be reached by more than one
geodesic.  Even more remarkably, there exists a set of endpoints
(lower branch of the thick pink curve in \fig{startphinf})
which are connected by both a null and a spacelike geodesic! Also,
as a corollary, it is no longer true that the endpoints of null
geodesics lie on the boundary of the set of endpoints of spacelike
geodesics. Correspondingly, the enpoints of the null geodesics reflect this
behaviour, as can be seen easily in \fig{startphrho}. While
$\Delta \ph$ slowly increases with increasing $\ro$, the
endpoints never approach a straight line, as would be the case for
the black hole geometry. This in turn means that the endpoints
are less sensitive indicator of the star geometry than in
the black hole case.

%

\subsection{Comparison between star and black hole }

Let us now ask how does the geometry vary with $\ro$.  Clearly, as $\ro \to 0$, the geometry becomes that of pure AdS; hence the star's utility in studying small perturbations on AdS.
One might naively expect that in the opposite limit, as we make the internal density very large, the star should start behaving more and more like a black hole.  Unfortunately, this is not the case, as will be explained below.

First of all, to compare the star's geometry with that of a \SAdS\ black hole, the most sensible map of parameters is to identify the black hole's mass with the star's total mass, so that the asymptotic geometry matches.  In other words, for the black hole, we use the metric of the form in \starmet, but with
\eqn{metBH}{
f(r) = h^{-1}(r) =  {r^2 \over R^2} + 1 - {M \over r^2} \ , \qquad {\rm where}
\ \ M \equiv \lim_{r \to \infty} m(r)
}
%
\begin{figure}[htbp]
\begin{center}
\includegraphics[width=6in]{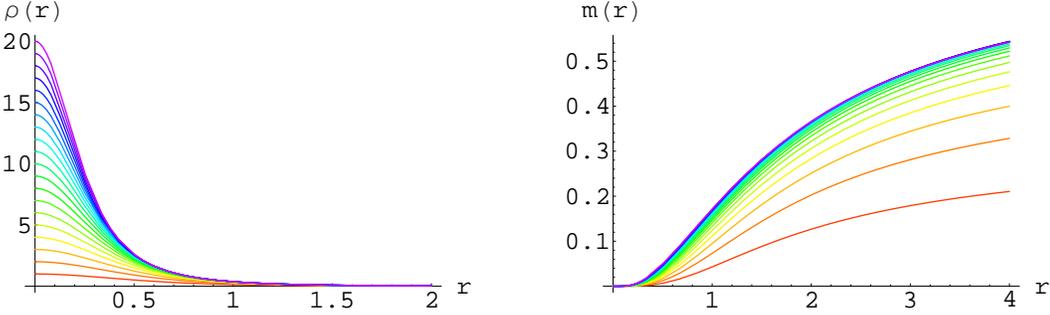}
\caption{Density profile, and corresponding mass profile, for a star with various values of $\ro$ (which can be read off from values of $\rho(r=0)$).}
\label{starmrho}
\end{center}
\end{figure}
Now, consider the variation of the star's density and mass profiles as we increase $\ro$.  This is plotted in \fig{starmrho}.  We see that as the star's total ``extent" does not seem to change much with $\ro$ -- the star is always confined within a size of order the AdS radius.  This is of course due to the confining potential of the AdS geometry.   Now, if the total mass could be made arbitrarily large, then for $M \gtrsim 1$, the star would be confined within its own Schwarzschild radius, and should be viewed as a black hole.   However, we see that while the density can be increased, the mass is bounded from above, and becomes largely independent of $\ro$ after some point.
\begin{figure}[htbp]
\begin{center}
\includegraphics[width=6in]{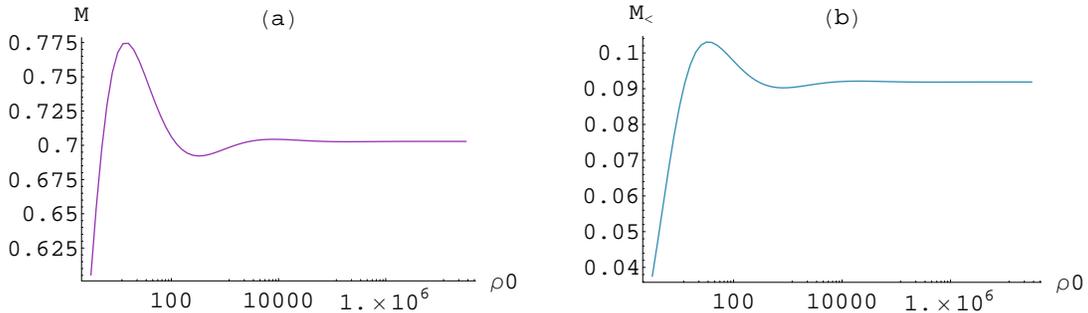}
\caption{Star's total mass (a), and the fraction of the star's mass confined to within this radius, plotted as a function of $\ro$.}
\label{starMrho}
\end{center}
\end{figure}
This is illustrated more clearly in \fig{starMrho}, where (a) the total mass, and (b) the fraction of the star's mass confined to within its effective Schwarzschild radius is plotted as a function of $\ro$.  The left plot shows clearly that the total mass is bounded\footnote{
In fact, this result is analogous to the similar result in the more familiar 4-dimensional, asymptotically flat static spherically symmetric spacetime, where given a fixed size $R_{\ast}$ of a star (with any equation of state), the maximum possible mass such a star can attain is $M_{\rm max} = 4 R_{\ast}/9$.  \cite{Wald:1984rg}
}
 from above by a rather small value: $M \le 0.775 R$.  Curiously enough, the mass is not a monotonic function of the internal density.\footnote{
This might correspond to onset of some sort of radial instability, though there does not seem to be any obvious pathology.  Correspondingly, we'll continue to include such high-$\ro$ solutions in our considerations. For the stars in 4-dimensional, asymptotically flat static spherically symmetric spacetime, \cite{Sorkin:1981jc} argued that an instability sets in at the turnover point where the mass function is non-monotone.
}
In fact, when examined in more detail, $M(\ro)$ appears to exhibit certain self-similarity; however we will not detour into this  intriguing observation further.
We also find that the effective Schwarzschild radius (\ie, what would be the Schwarzschild radius for a black hole with the total mass plotted in \fig{starMrho}a) is bounded by $r_+ \le 0.716 R$, so the star could at best be compared with a small \SAdS\ black hole.  Most importantly, \fig{starMrho}b shows that the mass contained within the effective Schwarzschild radius is a small fraction of the total mass (less than 11\%), so that the star does {\it not} approach a black hole in any regime.  In other words, we should not expect to obtain behaviour characteristic of a presence of an event horizon for any value of $\ro$.

\begin{figure}[htbp]
\begin{center}
\includegraphics[width=6in]{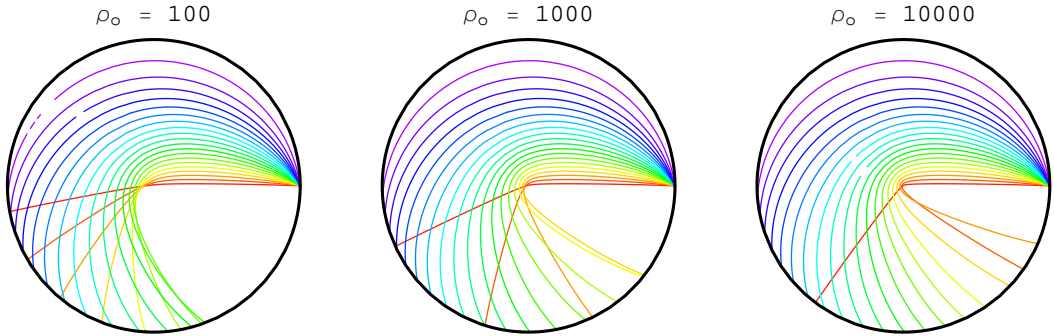}
\caption{Null geodesics in AdS star with three different values of $\ro$ as indicated, projected onto  a constant $t$ slice, for varying angular momentum to energy ratio.}
\label{starrphrho}
\end{center}
\end{figure}

We have seen that the star made up of a gas of radiation can never develop an event horizon, so it cannot look like a black hole, no matter how high its internal density $\ro$.  In \sec{statgeom} we have further argued for the absence of null circular orbits, and emphasized that the corresponding geodesic endpoints demonstrate a clear difference between a star and a black hole.  To illustrate this further, in  \fig{starrphrho} we plot the null geodesics, projected to a constant $t$ slice, for increasing values of $\ro$.  (This is the
same plot as in the left plot in \fig{stargeodsnull}, which had $\ro= 10$, redone for
$\ro= 100, 1000$, and $10000$, as indicated.)  We see that
even at very large $\ro$, there is no null orbit.  Even though the bending increases with $\ro$, it does so ever more slowly, rather analogously to the time delay for radial geodesics (\cf\ \fig{startdelay}).

\Appendix{Eternal black hole in AdS}
\label{BHspace}

We now discuss some interesting properties of spacelike geodesics
in the \SAdS\ background. The geodesic equations are given in
\geodbhs. The quantity of interest for spacelike geodesics is
$\(\D t (E,J),\D \ph (E,J)\)$ which depends on both $E$ and $J$.

\begin{figure}[t]
\begin{center}
\includegraphics[scale=0.8]{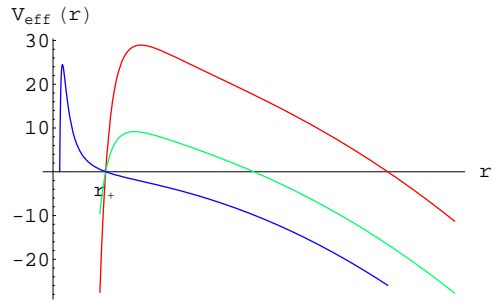}
\end{center}
\caption{Plots of $V_{eff} (r)$ for $J=5,3,0.1$, which correspond
to red, green and blue curves respectively. The horizon is at
$\rh=.78$~(for $\mu=1$). For $J < \rh$ (e.g. blue curve),
geodesics will fall into the horizon and will not come back to the
same boundary for any $E$. Similarly for $E > E_{max}$. $E_{max}$
is the maximal value of $V_{eff}$.} \label{vpot}
\end{figure}

The effective potential $V_{eff}(r)$ given in \bheffv\ for spacelike geodesics has
two real zeros at $r=J$ and $r = \rh$, respectively. For $J <
\rh$, $V_{eff}(r) <0$ for $r>\rh$, so for any $E$, the spacelike
geodesic will fall into the horizon\footnote{Subsequently it will
either reach the other boundary or fall to the singularity.} and
will not come back to the same boundary. We are not interested in
these geodesics.  For $J>\rh$, $V_{eff}(r)$ has a maximum $V_c (J)
= V_{eff}(r_m)$ at a value $r_m>\rh$ (see \fig{vpot}). For
$E^2>  E_{max}^2 (J) = V_c (J)$, the geodesic will again fall into
the horizon. Thus for a spacelike geodesic to come back to the
same boundary, $J$ and $E$ should lie in the range $J \in
(\rh, \infty)$ and $E \in (0, E_{max} (J))$ for a given $J$.

\begin{figure}[t]
\begin{center}
\includegraphics[scale=1]{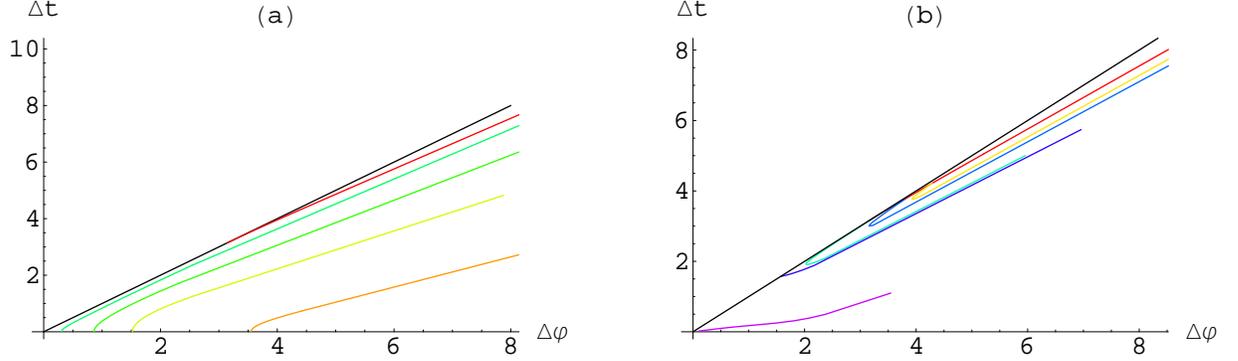}
\end{center}
\caption{The end points of various spacelike geodesics in AdS BH
(with $\mu=1$ and $\rh =0.78$). Black and red line correspond to
boundary and bulk null geodesics respectively as in \fig{yp}. (a):
Various curves show how end points change as $E$ is varied from
$E=0$ to $E_{max} (J)$ with fixed $J$ corresponding to $J =0.8,
1.2,\,2.2,\,6.5$. The curve with larger $J$ is farther away from
the the lines corresponding to the null geodesics. Note that at
$E=0$, all lines start at real axis and as $E \to E_{max} (q)$,
$\Delta t$ and $ \Delta \ph$ go to infinity. (b): Various lines
describe how end points change with $J$ for a fixed $\al=E/J=1.1,\
1.08, \ 1.02, \ 1, \ 0.2 $. The line with bigger value of $\al$ is
closer to the lines of null geodesics. $J$ varies from some finite
$J_c$ to $\infty$. It is clear that as $J \to \infty$ all curves
approach their corresponding points on the null curve. Note that a
curve with $0<\al < 1$ will end up at the origin as $J \to
\infty$. A curve with $\al =1$ could end up anywhere in the
diagonal line between the origin and $(\pi,\pi)$ depending the
value of $E-J$ which should be fixed in the limit $E,J \to
\infty$. } \label{qplot}
\end{figure}

For a fixed $J > \rh$, at $E=0$, $(\D \ph (0,J), \D t (0,J)) = (\D
\ph_c (J), 0)$. $\D \ph_c (J)$ is a monotonic function of $J$. As
$J \to \infty$, $\D \ph_c (J) \to {2 \ov J} \to 0$, while as $J
\to \rh$, $\D \ph_c (J) \to \infty$ logarithmically. As $E$ varies
from $0$ to $E_{max} (J)$, $\D t (E,J)$ and $\D \ph (E,J)$ appear
to monotonically increase to infinity (see \fig{qplot}). As $E \to
E_{max} (J)$, it can be shown that
 \be
 \D t (E, J) \approx {1 \ov s (J)} \,  \D \ph (E,J) \to \infty
 \ee
 where $s(J)$ is given by
 \be \label{rjh}
 s (J) = {J \, f(r_m) \ov E_{max} \, r_m^2} = \sqrt{1 + {\mu \ov r_m^4
 (J)}}
 \ee
Recall that $r_m (J)$ is the maximum of $V_J (r)$. Note that $r_m
(J=\rh) = \rh$ and as $J \to \infty$, $r_m (J) \to (2 \mu)^\ha$.
One can also check numerically that $r_m (J)$ is a monotonic
function of $J$. From the second equality of (\ref{rjh}), $s(J)$
is then a  monotonic function of $J$ decreasing from $s(J=\rh)
=\sqrt{2 + {1 \ov \rh^2}}$ to $s(J=\infty)= \am$, where $\am$ was
introduced in (\ref{deal}).

From the above discussion we conclude that the end points of
spacelike geodesics cover the whole region bounded by the red
curve corresponding to the bulk null geodesics, the straight line
from the origin to $(\pi,\pi)$, and the horizonal axis in
\fig{yp}. It is important to emphasize that $\ph$ is a
periodic variable, so in various figures, the fundamental region
is the strip  $\ph \in (-\pi, \pi]$. As a result we conclude that

\begin{enumerate}
\item For any end point $(\D \ph,\D t)$ of a null geodesic (including
boundary ones), there are an infinite number of spacelike
geodesics ending at that point due to the periodicity of $\ph$.
\item The bulk and boundary null geodesics can also share their
end points for special values of $\D t$ and $\D \ph$.
\end{enumerate}

Finally let us comment on how to obtain null geodesics from
spacelike geodesics by taking a limit. It is easy to see that we
see that the equations for spacelike geodesics approach those of
null geodesics in the limit of $E, J \to \infty$ while keeping the
ratio $\al=E/J$ fixed. More explicitly, the end point of a
spacelike geodesic with $0 \leq \al<1$  will approach the origin
in the $J \to \infty$ limit. A spacelike geodesic with $\al =1$
will approach a boundary null geodesic in the limit. Its end point
can be anywhere in the straight line between the origin and
$(\pi,\pi)$ depending the value of $E-J$ which should be fixed in
the limit $E,J \to \infty$.  A spacelike geodesic with $\al \in
(1,\am)$ will approach a bulk null geodesic in the limit. A
spacelike geodesic with $\al >\am$ will not come back to the same
boundary. Fig.~\ref{qplot} plots the end points of spacelike
geodesics with $E/J$ fixed.

\Appendix{No coming back for Poincar\'e patch}
\label{apppoincare}

In the following we argue that for field theories formulated on $\R^{d-1,1}$ and states respecting the full Poincar\'e symmetry there
are no new light-cone singularities. As discussed in \sec{nullgsing}, the presence of
light-cone singularities in the boundary field theory is governed by the properties of bulk
null geodesics. For field theories formulated on  $\R^{d-1,1}$ we will show that there
are no null geodesics through the bulk spacetime connecting boundary points. This implies
that the only null geodesics that connect points on the boundary are those that lie entirely
within the boundary and hence the nature of singularities in the correlation function
is entirely determined by the boundary causal structure.

To establish the absence of null geodesics connecting boundary points through the bulk, let
us consider a bulk spacetime with negative cosmological constant foliated by $\R^{d-1,1}$ slices.
The metric can be written in the warped-product form by picking a metric $g_{mn}(x^i)$ on the boundary and a radial coordinate $r$:
\eqn{poincaremet}{
ds^2 = e^{2 A(r)} \, g_{mn} \, dx^m \, dx^n + dr^2
}
For this geometry one can write down the geodesic equations as
\eqn{poingeods}{\eqalign{
\dot{r}^2 + e^{2 A(r)} \, g_{mn} \, \dot{x}^m \dot{x}^n &= \kappa \ ,\cr
\ddot{r} - A'(r) \, e^{2 A(r)} \, g_{mn} \, \dot{x}^m \dot{x}^n &= 0\,  }}
with $\kappa$ being $\pm,0$ for spacelike, timelike and null geodesics, respectively.
By eliminating the boundary directions we can write an effective classical particle in a potential equation for the motion in the radial direction:
\eqn{poingeodsb}{
\ddot{r} + A'(r) \, (\dot{r}^2 - \kappa) = 0}

The issue of whether geodesics emanating from the boundary into the bulk turn around, can
now be analyzed simply in the effective problem for the radial motion \poingeodsb. For the
case of null geodesics which we are especially interested in the first integral from \poingeods\
gives:
\eqn{poinfirst}{
\dot{r} = {C \over e^{A(r)} } \, }
implying that $A(r) \to \infty$  for the geodesic to  turn around (since $\dot{r} =0$ by definition
at the turn around point).

However, $A(r)$ cannot diverge to $\infty$ -- the energy condition on the stress tensor require that $A'(r)$ be a monotonically decreasing function \cite{Freedman:1999gp}. This effectively implies that $A(r)$ decreases from $\infty$ at the AdS boundary where the field theory is formulated. Physically divergence of $A(r)$ would look like we have second timelike boundary in the geometry which again is disallowed by the null geodesic convergence condition in spacetimes where the matter sources respect the null energy condition.

Note that this result is contingent on the Poincar\'e symmetries of the boundary  being preserved.\footnote{We thank Simon Ross for pointing this out to us and suggesting the example described.} We can have geodesics that return to the boundary even in the non-compact case when we consider states that break the boundary Poincar\'e invariance. A simple example is to consider the global AdS-Schwarzschild geometry that is sliced in the Poincar\'e coordinates (\cf, \cite{Horowitz:1999gf} for a boundary description of this state). Here it is possible to exploit the black hole, which breaks the Poincar\'e symmetry to slingshot oneself back to the boundary in a timescale shorter than the AdS time.

\Appendix{Collapsing shell in AdS}
\label{timede}

In \sec{horform} we derived a scaling formula for
$t_o$ as a function of $\delta t  = t_h -t_i$ using a ray tracing calculation. The exponential relation is given in (\ref{delay}) and intuitively we can imagine this arising from the red-shift in the vicinity of the horizon.  We now proceed to derive this result more explicitly, by finding actual radial null geodesics in the collapse background.  In this Appendix we constrain ourselves to the sharp shell cleanly separating AdS and \SAdS; in \App{vaidyaapp} we consider the more general smeared shell collapse modeled by a Vaidya spacetime and also extend our analysis to non-radial geodesics.

Consider the set-up as sketched in \fig{collapsePD}.  There are two interesting limits to consider as discussed in \sec{bhsing}.
In particular, we wish to obtain the leading behaviour both as $t_i \to t_h^{\pm}$ and as $t_i \to t_s$, as indicated in \fig{collapsePDs}a and b, respectively.
In fact, in order to derive both (\ref{delay}) and the scaling exponent $\ga$ introduced in \sec{bhsing}, we will find the full exact expression for $t_o$ in terms of $t_i$ and the parameters $t_s$ and $\rh$ describing the shell.
It is interesting to examine the geodesic behavior both in $d=3$ and in $d=5$. The distinction of course is engendered by the fact that the BTZ black hole in $d=3$ being an orbifold of AdS$_3$ is a simpler geometry.  On the other hand, the physics of the type III geodesics of \fig{collapsePD}b is different is quite different since spacelike geodesics do not bounce off the BTZ singularity, so we can meaningfully consider only the $t_i \to t_h^{-}$ scaling.

The strategy for calculating radial null geodesics will be to write the geodesics in the different coordinate patches (such that in each patch we can use the metric of the form in (\ref{metricgen})), and then patch them together using the fact that both the geodesics and the shell are null.  The three regions of interest are depicted in \fig{collapsePDs}c, and correspond to
\begin{enumerate}
\item pure AdS (before/inside the shell)
\item \SAdS, outside the horizon (outside the shell)
\item \SAdS, inside the horizon (but still outside the shell)
\end{enumerate}
\subsection{$d=3$}

As a warm-up, let us first focus on $d=3$. For simplicity, we use
the metric (\ref{metricgen}) with $f(r) = r^2 + 1$ in region 1,
and $f(r) = r^2 - r_+^2$ in regions 2 and 3. We will write the
expressions for geodesics in 3 distinct regions (even though in 3
dimensions, region 3 is not so relevant since radial spacelike
geodesics in BTZ do not bounce off the singularity).  Calculating
the geodesics as before, we have for pure AdS: \eqn{}{ t_{AdS} (r)
= t_0 \pm \( \tan^{-1} r - \tan^{-1} r_0 \)} and for BTZ:
\begin{eqnarray}
t_{BTZ} (r) &= t_0 \mp {1 \over r_+} \( \tanh^{-1} {r \over r_+} -
\tanh^{-1} {r_0 \over r_+} \) \qquad &{\rm inside} \cr &= t_0 \pm
{1 \over 2 r_+} \ln \[ \( {r -  r_+ \over r + r_+} \)  \( {r_0 +
r_+ \over r_0 - r_+} \) \] \qquad &{\rm outside}
\end{eqnarray}
where we denote the initial conditions by $t(r_0) \equiv t_0$. We
can now patch these together to see when a geodesic starting at
$t=t_i, r = \infty$ reemmerges back out to $r = \infty$.  Denoting
this time $t_o$, we can express $t_o$ in terms of $t_i$, $t_s$,
and $r_+$, obtaining: \eqn{toBTZ}{  t_o =  t_s + {1 \over r_+} \ln
\left( { \tan( {t_s-t_i \over 2} ) + r_+ \over \tan( {t_s-t_i
\over 2} ) - r_+ } \right) \qquad {\rm if} \ \ t_o < t_h} Matching
where the shell intersects the horizon, we can relate $t_s$ to
$t_h$ as follows: \eqn{tsBTZ}{t_s = t_h +2 \, \tan^{-1} r_+ }
Finally, substituting (\ref{tsBTZ}) into (\ref{toBTZ}) and
expanding for small $\d t$, we obtain the scaling behaviour
\eqn{scalingBTZ}{  t_o \sim -{1 \over r_+} \, \ln \abs{\d t}
\thuss \d t \sim  -  \, e^{-r_+ \, t_o} = -  \, e^{-\kappa \, t_o}
\qquad {\rm as} \ \ t_o \to \infty } where $\k = r_+$ is the
surface gravity of the BTZ black hole.

Unfortunately, in this set-up we can only answer the first part of the questions raised in \sec{bhsing}: as $\d t \to 0^-$, $t_o $ diverges logarithmically with the coefficient given by the black hole temperature, as one would naively expect.  To see the other scalings, $t_i \to t_h^+$ and 
$t_i \to t_s$, discussed in \sec{bhsing} , we need to go to higher dimensions.\footnote{ If we nevertheless ignore the fact that spacelike geodesics do not bounce off the singularity, and
consider the two null geodesics meeting at the singularity, then we would expect to find 1) the same scaling behaviour (same coefficient) as in (\ref{scalingBTZ}) for $t_i > t_h$, and 2) $(t_s-t_i) \sim (t_o - t_s)$ (\ie\ $\gamma = 1$).   }

\subsection{$d=5$}

Now let us consider $d=5$. Whereas the time-radius relation for
radial geodesics is independent of the dimension in pure AdS, we
have a more complicated relation $t(r)$ for the black hole
geometry than in the BTZ case. Writing
\eqn{}{ t(r) = t(r_0) \pm
\int_{r_o}^{r} {d \bar{r} \over f(\bar{r}) } }
and rewriting the metric function outside the shell as
\eqn{}{ f(r) = { (r^2 +
\rho_+^2) \, (r^2 - r_+^2) \over r^2 } }
where $ \rho_+^2 \equiv 1 + r_+^2$, we can express the indefinite integral pertaining to
outside and inside horizon regions respectively as
\eqn{}{ \int
{dr \over f(r)} = {1 \over r_+ \, \k} \( \rho_+ \, \tan^{-1} {r
\over \rho_+} - r_+ \, \tanh^{-1} {r_+ \over r} \) \qquad {\rm
for} \ \ r > r_+ }
\eqn{}{ \int {dr \over f(r)} = {1 \over r_+ \,
\k} \( \rho_+ \, \tan^{-1} {r \over \rho_+} - r_+ \, \tanh^{-1} {r
\over r_+} \) \qquad {\rm for} \ \ r < r_+ }
where $\kappa = { 1+2r_+^2 \over   r_+} $ is the surface gravity of the black
hole. For a geodesic crossing the horizon, the infinite
contribution from the coordinate singularity cancels out, so if
$r_1>r_+$ and $r_2<r_+$, say, we would obtain
\eqn{}{t_{SAdS}(r_2)
= t_{SAdS}(r_1)- {1 \over r_+ \, \k} \( \rho_+ \, \tan^{-1} {r
\over \rho_+} \biggr|^{r_2}_{r_1} + r_+ \, \tanh^{-1} {r_+ \over r_1}
- r_+ \, \tanh^{-1} {r_2 \over r_+} \)  }

Applying this to the geodesics of type $II$ and $III$ in
\fig{collapsePDs}a, we obtain  the full exact expression for
$t_o$ in terms of  $t_s$, the parameters specifying the black
hole, and the radius of where the shell crosses our geodesic
$r_c$:
\eqn{toII}{ t_o = t_s + {2 \over r_+ \, \k} \( \rho_+ \,
{\pi \over 2}
 - \rho_+ \, \tan^{-1} {r_c \over \rho_+} +
r_+ \, \tanh^{-1} {r_c \over r_+} \)  \qquad {\rm for \ geod. \ }
II }
\eqn{toIII}{ t_o = t_s + {2 \over r_+ \, \k} \( - \rho_+ \,
\tan^{-1} {r_c \over \rho_+} + r_+ \, \tanh^{-1} {r_c \over r_+}
\)  \qquad {\rm for \ geod. \ } III }
 where (in both cases) the crossing radius is given by
 \eqn{}{ r_c =  \tan \( {t_s - t_i \over 2} \) }
 and the time of the shell is related to the time of horizon formation by
\eqn{}{ t_s = t_h + 2 \, \tan^{-1} \, r_+   }

We can easily expand this out for small $\d t$, to check that in
both cases
\eqn{}{ \abs{\d t} \,  \approx  \, e^{-\kappa \, t_o} \thuss
t_o \sim {1 \over \kappa} \, \ln {1 \over \abs{\d t}} \ . }
 This is
again consistent with our expectations (\ref{delay}), and indeed of the same
form as the result obtained for the 3-dimensional case.

Finally, we can also use expression (\ref{toIII}) to consider the
case of  \fig{collapsePDs}b. In particular, if we let $t_s
-t_i \equiv \eps$, then expanding $t_o - t_s$ to third order in
$\eps$, we find that
\eqn{gammascaling}{t_o - t_s \approx {1 \over 12 \,
r_+^2 \, \rho_+^2 } \, \eps^3 \ . }
This in particular implies that the scaling parameter $\ga$ introduced in \sec{bhsing} is $3$  for black hole formation in $d=5$.
Intriguingly, we can generalize this to $d$ dimentions:  For $d$-dimensional collapse spacetime, $\ga = d-2$.

\Appendix{Smeared shell in AdS and non-radial geodesics}
\label{vaidyaapp}

 In \App{timede}, we have considered radial null geodesics in the collapsing null shell geometry.  In particular, we calculated $t_o$ as a function of $t_i$ and the shell parameters.  As discussed in \sec{horform}, being able to read off the value of $t_i$ for which $t_o$ diverges from the gauge theory correlators, we can automatically determine the horizon formation time $t_h$ ($\thus t_H$).
 Here we wish to extend this analysis in two directions.  First, we want to consider a more general (and physically more realistic) background corresponding to a smeared shell.  Second, we want to analyze non-radial geodesics, as motivated in \sec{horform}.  To that end, we will discuss general null geodesics in Vaidya-AdS spacetime.

\subsection{Vaidya for AdS}

Consider the stress tensor for a spherical null gas
 \be T_{vv} =
h (r,v)
 \ee
The metric can be written as
 \be \label{vaidya}
ds^2 = - f(r,v) \, dv^2 + 2 \, dv \, dr + r^2 \, d \Om^2_3
 \ee
with
 \be
 f(r,v) = r^2 +1 - {m(v) \ov r^2}
 \ee
The covariant conservation $\nabla^\mu T_{\mu \nu} = 0$ of $T$
implies that\footnote{Note that $\Gamma^{v}_{rv}=\Gamma^{r}_{rr} =
0$.}
 \be
 h(r,v) = {1 \ov r^3} q (v)
 \ee
The Einstein equations now reduce to (with $\Lambda = -6$)
 \be
 m'(v) = {2 \ov 3} q (v)
 \ee
For the sharp shell we take $q(v)$ has the form of a delta function, \ie,
 \bea
 m (v) & = &  0 \qquad v< v_0 \\
      & = & \mu \qquad v > v_0
      \eea
      with $q (v) = {3 \ov 2} \mu \, \delta (v-v_0)$.
Note that in this case the $v$-dependence disappears before and after the shell; the spacetime (\ref{vaidya}) is static in both regions (but not globally static because of the shell), written in ingoing coordinates,
 \be \label{nuco}
 v = t + z , \qquad dz = {dr \ov f}
\ee

\subsection{Null geodesics}

Let us now consider null geodesics in the background
(\ref{vaidya}).
Writing the second-order geodesic equation (in the equatorial plane) directly yields
\begin{eqnarray}
& \ddot{v} + \half \, \p_r f(r,v) \, \dot{v}^2 - r \, \dot{\ph}^2 = 0 \cr
& \ddot{r} - \half \,
\[\p_v f(r,v) + f(r,v) \, \p_r f(r,v) \] \, \dot{v}^2  - r \, f(r,v)  \, \dot{\ph}^2= 0 \cr
& \ddot{\ph} + {2 \over r} \, \dot{r} \, \dot{\ph} = 0
\end{eqnarray}
For any given $f(r,v)$, we can solve these numerically to find any geodesic through the bulk.  The function  $f(r,v)$ with which we choose to model the smeared shell in \sec{horform} is
\eqn{vaidyfa}{
f(r,v) = r^2 + 1 - {\mu \over r^2} \( {1+ \tanh{v \over v_s} \over 2} \)
}
which smoothly interpolates between AdS at $v \to - \infty$ and \SAdS\ as $v \to + \infty$. In particular, the shell is inserted at\footnote{
Since both AdS and \SAdS\ geometries are static, we can WLOG set the time $t_s$ of the shell's creation on the boundary; for convenience we choose $t_s = -\pi/2$.
} $v=0=t_s + \pi/2$, and has `thickness' $v_s$.  This provides a convenient regulator of the thin shell collapse, since as $v_s \to 0$, we recover the collapse spacetime (\ref{metricgen}) written in ingoing coordinates.

However, to find the geodesics in the collapse spacetime, and specifically to determine $t_h(\a)$, we can use a simpler and more elegant method, which we now indicate.
The geodesics can be obtained from the following action
 \be
 S = \int d \tau \, \(-\ha f \dot v^2 + \dot v \dot r + \ha r^2
 \dot \ph^2 \)
 \ee
The canonical momenta are
 \bea
 J = r^2 \dot \phi, \qquad E = - {\p L \ov \p \dot v} = - \dot r + f
 \dot v, \qquad {\p L \ov \p \dot r} = \dot v
 \eea
Note that $J$ is conserved, while $E$ is in general not. Note that
$J$ can be scaled to be $1$ by a rescaling of $\tau$.
 As before, we will denote $\al = {E \ov J}$ and set $J=1$.
 Another
first integral of the system is given by
 \be \label{rdotsquared}
 \dot r^2 + {f  \ov r^2} = \al^2
 \ee
  The equations of motion can now
 be written as
 \bea
  \dot \al & = &  \ha {\p f \ov \p v} \dot v^2 \\
  \ddot v & = & -\ha {\p f \ov \p r} \dot v^2  + {1 \ov r^3}
 \eea
From the above equations we can also derive
 \be
 \ddot r = \ha {\p f \ov \p v} \dot v^2- {1 \ov 2r^2} {\p f \ov \p r}
 + {f \ov r^3}
 \ee
where we have used that
 \be \label{vdot}
  \dot v = {\al + \dot r \ov f}
  \ee

Now we apply the above equations to the Vaidya spacetime with a sharp shell, in which
case
 \be
 {\p f \ov \p v} = - {\mu \ov r^{d-3}} \, \delta (v-v_0)
 \ee
The discontinuity of $f$ across $v=v_0$ is thus given by $\delta f
= - {\mu \ov r^{d-3}}$. One can readily conclude from the above
equations that both $\al$ and $\dot r$ jump across $v=v_0$, in
fact by the same amount, while $\dot v$ is continuous. More
explicitly we find that
 \be \label{shift}
 \delta \dot r= \delta \al = - {\mu \ov 2 r^{d-3}} \, \dot v \biggr|_{v_0}
  < 0
 \ee

Let us now return to the problem of determining $t_h(\a)$ for the collapse geometry.  Realizing that $\a$ jumps across the shell, let us work backwards by considering what feature of the geodesics makes $t_o$ large.  The relevant spacetime to consider is \SAdS, wherein the geodesics follow from the potential (\ref{bheffv}) drawn in \fig{vpot1}.  Since the top of the potential is always outside the horizon (as follows from the discussion in \sec{bhthermal}, $r_m^2=2\mu = 2 \rh^2 \, (\rh^2 + 1) > \rh^2 $), the only way that a non-radial null geodesic could emerge at infinite time would be to get trapped in the unstable circular orbit at the top of the effective potential.  This requires
\eqn{alphaoVeff}{
\a_o^2  = V_{eff}(r_m) = 1 + {1 \over 4 \mu}
}
Knowing the $\a$ in the \SAdS\ part of the spacetime, we can now use the expression (\ref{shift}) for the jump in $\a$ across the shell, to find what initial conditions (\ie\ what $\a_i$ in AdS) we need to start with to achieve this $\a_o$.
Using (\ref{vdot}), (\ref{rdotsquared}), and the expression for $f(r)$ in AdS (\ref{fin}), all evaluated at some crossing radius $r_x$ where the geodesic intersects the shell, we can reexpress (\ref{shift}) as follows:
\eqn{alphajumprx}{
\alpha_o = \alpha_i
      - {\mu \over 2 r_x^2} \,  { \alpha_i + \sqrt{\alpha_i^2 - {r_x^2 + 1 \over r_x}} \over r_x^2 + 1}
}
Finally, we can express the crossing radius $r_x$ in terms of the initial time $t_i$ by following the geodesic with parameter $\a_i$ from $r_x$ back to its starting point on the boundary at $t_i$.  This is given by
\eqn{rx}{
 t_s - t_i = \tan^{-1} r_x + \tan^{-1} {\sqrt{(\alpha_i^2 - 1) \, r_x^2 -1 } \over \alpha_i }
}
Note that $t_i$ should be viewed as a function of $\a_i$, and for $t_o \to \infty$, this is nothing but $t_h(\a)$ of \sec{horform}, where we have been using $\a \equiv \a_i$.

Hence, to find $t_h(\a)$, we can solve the system of equations (\ref{alphaoVeff}), (\ref{alphajumprx}), and (\ref{rx}).  We first solve (\ref{alphaoVeff}) and (\ref{alphajumprx}) for the crossing radius $r_x(\a_i)$, and then substitute this into  (\ref{rx}) to find $t_h(\a) = t_i$, as a function of $\a_i \equiv \a$ and the shell parameters, $\mu$ and $t_s$.
This determines the black curve plotted in \fig{vadtj}.
 Note that the radial geodesic limit ($\a \to \infty$) is continuous, despite the fact that only the radial geodesic can truly sample the horizon formation event, since only the null outgoing radial geodesics can escape from the close vicinity of the horizon.


\bibliographystyle{utphys}
\providecommand{\href}[2]{#2}\begingroup\raggedright\endgroup


\end{document}